

\input eplain

\def\toprule{\vskip1.5pt\hrule height0.8pt\vskip1.5pt}
\def\midrule{\vskip1.5pt\hrule\vskip1.5pt}
\def\bottomrule{\vskip1.5pt\hrule height0.8pt\vskip1.5pt}

\def\nomenclature[#1]#2#3{\line{{#2\hfill} \hbox to0.85\hsize {#3\hfill}}}

\newcount\fignumber
\def\figdef#1{\global\advance\fignumber by 1 \definexref{#1}{\number\fignumber}{figure}\ref{#1}}
\def\figdefn#1{\global\advance\fignumber by 1 \definexref{#1}{\number\fignumber}{figure}}
\let\figref=\ref
\let\figrefn=\refn
\let\figrefs=\refs

\newcount\tabnumber
\def\tabdef#1{\global\advance\tabnumber by 1 \definexref{#1}{\number\tabnumber}{table}\ref{#1}}
\def\tabdefn#1{\global\advance\tabnumber by 1 \definexref{#1}{\number\tabnumber}{table}}
\let\tabref=\ref
\let\tabrefn=\refn
\let\tabrefs=\refs

%
\ifx\pdfoutput\undefined
\input epsf

\def\figscale#1#2{\epsfxsize=#2\epsfbox{#1.eps}}
%
\else

\def\figscale#1#2{\pdfximage width#2 {#1.pdf}\pdfrefximage\pdflastximage}
\fi


\newcount\scount \scount=0
\newcount\sscount \sscount=0




\makeatletter
\def\section#1\par{
  \vskip\z@ plus.3\vsize\penalty-250
  \vskip\z@ plus-.3\vsize\bigskip\vskip\parskip
  \global\advance\scount by1
  \sscount=0
  \writenumberedtocentry{section}#1{\the\scount}
  \definexref#1{\the\scount}{section}
  \message{#1}
  \noindent\the\scount.\quad{\bf #1}\nobreak\smallskip\noindent}
\makeatother

\def\subsection#1{
  \global\advance\sscount by1
  \smallskip
  \noindent{~~\the\scount.\the\sscount~~{\bf{#1.~}}}}

%


\centerline{\bf{Natural Convection Heat Transfer From an Isothermal Plate}}
\medskip
\centerline{Aubrey G. Jaffer}
\centerline{e-mail: agj@alum.mit.edu}

\beginsection{Abstract}


\def\Nuz{{N\!u_0}}
\def\Nuzq{{N\!u'_0}}
\def\Nuzs{{N\!u^*_0}}
\def\Nuol{\overline{N\!u}}
\def\Nuols{\overline{N\!u^*}}
\def\Nuolq{\overline{N\!u'}}
\def\Pra{{Pr}}
\def\Ra{{Ra}}
\def\Gr{{Gr}}
\def\Rey{{Re}}
\def\diff{{\rm d}}
\def\Ls{{L\!^*}}
\def\Lsw{{L\!^*\!_w}}



{\narrower

  Using boundary-layer theory, natural convection heat transfer
  formulas which are accurate over a wide range of Rayleigh numbers ($Ra$) were
  developed in the 1970s and 1980s for vertical and downward-facing
  plates.  A comprehensive formula for upward-facing plates remained
  unsolved because they do not form conventional boundary-layers.

  From the thermodynamic constraints on heat-engine efficiency, the
  novel approach presented here derives formulas for natural
  convection heat transfer from isothermal plates.
  The union of four peer-reviewed data-sets spanning $1<Ra<10^{12}$
  has 5.4\% root-mean-squared relative error (RMSRE) from the new
  upward-facing heat transfer formula.

  Applied to downward-facing plates, this novel approach outperforms
  the Schulenberg (1985) formula's 4.6\% RMSRE with 3.8\% on four
  peer-reviewed data-sets spanning $10^6<Ra<10^{12}$.

  The introduction of the harmonic mean as the characteristic-length
  metric for vertical and downward-facing plates extends those
  rectangular plate formulas to other convex shapes, achieving 3.8\%
  RMSRE on vertical disk convection from Hassani and Hollands (1987)
  and 3.2\% from Kobus and Wedekind (1995).

\par}



\bigskip
 This research did not receive any specific grant from funding
 agencies in the public, commercial, or not-for-profit sectors.

\beginsection{Table of Contents}

\readtocfile


\section{Introduction}

  Natural convection is the flow caused by nonuniform density in a
  fluid.  It is a fundamental process with applications from
  engineering to geophysics.

  When a stationary, immersed object changes temperature, nearby fluid
  can change density as it warms or cools.  Under the influence of
  gravity, density changes cause fluid to flow.  The rates of fluid
  flow and heat transfer from the object grow until reaching a
  plateau.
  This investigation seeks to predict the overall steady-state heat
  transfer rate from an external, flat, isothermal surface inclined at
  any angle in a Newtonian fluid.

  An ``external'' plate is one that fluid can flow around freely,
  especially horizontally.  If enclosed, the enclosure must have
  dimensions much larger than the heated or cooled surface.
  Natural convection in an enclosure of size comparable to the heated
  or cooled surface can organize into cells of Rayleigh-B\'{e}nard
  convection, which is not treated here.

  The characteristic-length $L$ is the length scale of a physical system.
  For many heat transfer processes, it is the volume-to-surface-area
  or area-to-perimeter ratio of the heated or cooled object.
  There are several characteristic-length metrics used for natural
  convection, some of which are valid only for convex objects.  This
  investigation focuses on flat plates with convex perimeters.


\subsection{Flow Topologies}
  There are three topologies of convective flow from external, convex
  plates.

  For a horizontal plate with heated upper face,
  streamlines photographs in Fujii and Imura~\cite{fujii1972natural}
  show natural convection pulling fluid horizontally from above the
  plate's perimeter into a rising central plume.  \figref{fig:above-flow},
  below, is a diagram of this upward-facing convection.
  Horizontal flow is nearly absent at the elevation of the dashed
  line.

  Kitamura, Mitsuishi, Suzuki, and Kimura~\cite{KITAMURA2015320} shows
  top-views of plumes from heated rectangular plates with aspect
  ratios between 1:1 and 8:1.  The plates with high aspect ratios have a
  plume over the plate's mid-line parallel to the longer sides, but
  not as long.

\medskip

  The streamlines photograph of a vertical plate in Fujii and
  Imura~\cite{fujii1972natural} shows fluid being pulled horizontally
  before rising into a plume along the vertical plate.
\medskip

  Modeled on a streamlines photograph in
  Aihara, Yamada, and End\"o~\cite{AIHARA19722535},
  \figref{fig:below-flow} is a flow diagram for a horizontal plate with
  heated lower face.  Unheated fluid below the plate flows
  horizontally inward.  It rises a short distance, flows outward
  closely below the plate, and flows upward upon reaching the
  plate edge.

\medskip
\vbox{\settabs 2\columns
\+\figscale{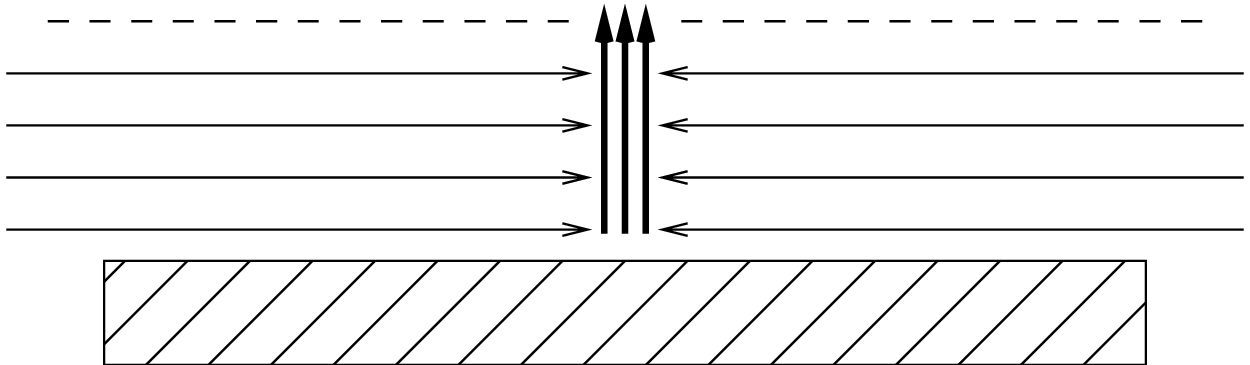}{230pt}&\hfill\figscale{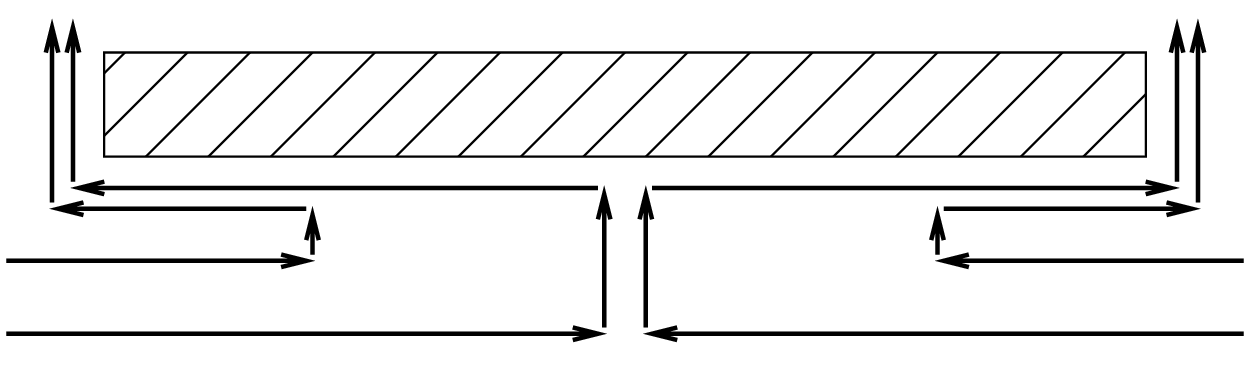}{230pt}&\cr
\+\hfil{\bf\figdef{fig:above-flow}\quad flow above a heated plate}&\hfil{\bf\figdef{fig:below-flow}\quad flow below a heated plate}&\cr
}
\medskip

  There is a symmetry in external natural convection; a cooled plate
  induces downward flow instead of upward flow.  Flow from a cooled
  upper face is the mirror image of flow from a heated lower face.
  Flow from a cooled lower face is the mirror image of flow from a
  heated upper face.

  Sublimation from an upper face is downward convection when the
  dissolved sublimate is denser than the fluid.  The rest of this
  investigation assumes a heated plate.

\medskip
  An important aspect of all three flow topologies is that fluid is
  pulled horizontally before being heated by the plate.  Pulling
  horizontally expends less energy than pulling vertically because
  the latter does work against the gravitational force.
  Inadequate horizontal clearance around a plate can obstruct flow
  and reduce convection and heat transfer.

\subsection{Fluid Mechanics}
  In fluid mechanics, the convective heat transfer rate is represented
  by the average Nusselt number~($\Nuol$).  The Rayleigh
  number~($\Ra$) is the impetus to flow due to temperature difference
  and gravity.  A fluid's Prandtl number~($\Pra$) is its momentum
  diffusivity per thermal diffusivity ratio.  These three ``variable
  groups'' are dimensionless (measurement units of the constituent
  variables cancel each other).

  The characteristic-length~$L$ scales~$\Nuol$; $\Ra$~is scaled
  by~$L^3$; $\Pra$~is independent of~$L$.


  Formulas for heat transfer can apply to mass transfer via analogous
  variable groups, such as Schmidt number~($Sc$)
  and~$\Pra$.  \figref{fig:hup} has Sherwood number~($\overline{Sh}$)
  instead of~$\Nuol$ on its vertical axis.

\subsection{Turbulence Versus Laminar Flow}
  Previous
  investigations~\cite{fujii1972natural,KITAMURA2015320,lloyd1974natural,churchill1975correlating}
  assumed that natural convection heat transfer formulas would differ
  substantially when the convection was turbulent versus laminar.  For
  their upward-facing plate, Lloyd and Moran~\cite{lloyd1974natural}
  reported that the transition from laminar to turbulent flow occurred
  at~$\Ra\approx8\times10^6$.  The straight line segments they fitted to their data at
  greater and lesser~$\Ra$ were disjoint at $\Ra=8\times10^6$.
  However, with their fit lines removed, if $\Ra\approx8\times10^6$
  represents a discontinuity, then it is one of several, and subsumed
  within the scatter of their measurements in \figref{fig:hup}.

\vbox{\settabs 1\columns
\+\hfil\figscale{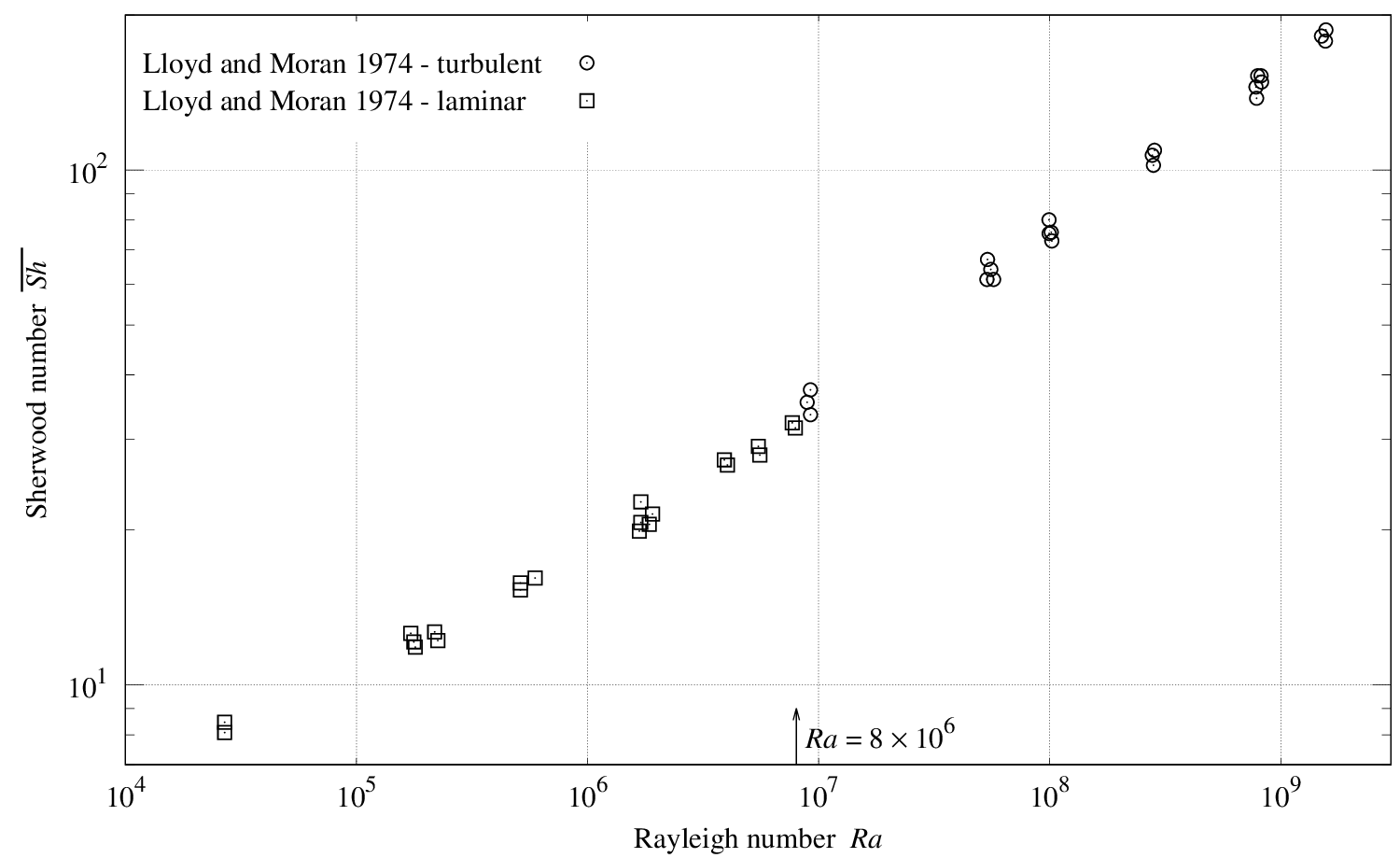}{468pt}&\cr
\+\hfil{\bf\figdef{fig:hup}\quad{upward convection heat transfer from horizontal plate}}&\cr
}

\smallskip
\noindent
  About their measurements of vertical and downward tilted plates,
  Fujii and Imura~\cite{fujii1972natural} wrote:

{\narrower
      ``Though the boundary layer was not always laminar near the
      trailing edge for large $\Gr\,\Pra$ [$=\Ra$] values, no influence of the
      flow regime on the data shown in [their] Fig.~6 is appreciable.''
\par}

\smallskip
\noindent
  Churchill and Chu~\cite{churchill1975correlating} concludes that one
  of its equations

{\narrower
  \noindent``\dots
  based on the model of Churchill and Usagi~\cite{AIC:AIC690180606}
  provides a good representation for the mean heat transfer for free
  convection from an isothermal vertical plate over a complete range
  of Ra and Pr from 0 to $\infty$ even though it fails to indicate a
  discrete transition from laminar to turbulent flow.''

\par}


\smallskip
  The lack of a significant transition between the rates of (mean)
  heat transfer in laminar and turbulent natural convection indicates
  that some more basic principle organizes natural convection.

\subsection{Thermodynamics}
  The fundamental laws of thermodynamics make no distinction between
  laminar and turbulent flows.  Considering a small object
  inside a very tall column of fluid as a closed system, natural
  convection is a heat-engine which converts the temperature
  difference between the object and fluid into flow of that fluid.
  The heated object is the heat source; fluid far above the object is
  the heat sink.

  System-wide thermodynamic constraints cannot be enforced locally,
  requiring a radical departure from boundary-layer analysis (which is
  based on local flow solutions of the Navier-Stokes equations).  This
  investigation solves algebraic equations in terms of average fluid
  velocity, heat conduction, and power flux.



\vfill\eject

\subsection{Not Empirical}
  Empirical theories derive their coefficients from measurements,
  inheriting the uncertainties from those measurements.  Theories
  developed from first principles derive their coefficients
  mathematically.  For example, the Blasius model of laminar flow
  coefficient $0.3320+$ is the solution of a differential equation
  (Lienhard and Lienhard~\cite{ahtt5e}).  Another example is the heat
  conduction shape factor for one face of a disk, which is exactly
  twice its diameter.
  The present theory derives from first principles; it is not
  empirical.\numberedfootnote{The conduction shape factor
  $q_{SS}^*=0.932$ is used in \ref{Vertical Rectangular Plate}.  It is
  derived exactly in \ref{Downward-Facing Circular Plate}.}  Each
  equation term is tied to an aspect of the plate geometry,
  orientation, and flow diagram.

  This investigation tests its theory on prior works' measurement data
  with as wide a range of $\Ra$ and $\Pra$ as possible.  Few of the
  cited studies provided estimated measurement uncertainties.
  Root-mean-squared relative error (RMSRE), introduced
  in \ref{Upward-Facing Measurements}, provides an objective,
  quantitative evaluation of each data set versus the theory.

\subsection{Detail}
  Were this theory derived using the usual fluid-mechanics techniques,
  derivation detail would be unnecessary.  However, this powerful new
  methodology is unlike those techniques.  The first derivations
  provide more detail to enable readers to adapt the present
  methodology to other fluid mechanics problems.

\smallskip
\centerline{\bf\tabdef{tab:sources}\quad{sources of measurements}}
\vbox{\settabs 10\columns
\toprule
\+ Source &&\quad Description && Fluid &\hfil $\Pra|Sc$ &\hfil Face & ~~$\Ra\ge$ & ~~$\Ra\le$ &\hfil $\pm$ &\cr
\midrule
\+ \cite{fujii1972natural} Fujii \& Imura &&\quad $5\times10$ cm && water &\hfil 5.0 &\hfil           up & $5.7\times10^6$ & $1.3\times10^{9}$ &\cr
\+ \cite{fujii1972natural} Fujii \& Imura &&\quad $30\times15$ cm && water &\hfil 5.0 &\hfil          up & $4.1\times10^9$ & $4.8\times10^{11}$ &\cr
\+ \cite{GOLDSTEIN19731025} Goldstein et al &&\quad sublimation && air &\hfil 2.50 &\hfil             up & $1.6\times10^0$ & $6.2\times10^{3}$ &\cr
\+ \cite{lloyd1974natural} Lloyd \& Moran &&\quad electrochemical && $\rm{H_2SO_4}$ &\hfil 2200 &\hfil up & $2.6\times10^4$ & $1.6\times10^9$ &\hfil 5\% &\cr
\midrule
\+ \cite{churchill1975correlating} Churchill \& Chu &&\quad Cheesewright && air &\hfil 0.70 &\hfil  vertical & $7.0\times10^3$ & $1.5\times10^{9}$ &\cr
\+ \cite{churchill1975correlating} Churchill \& Chu &&\quad Jakob && air &\hfil 0.70 &\hfil         vertical & $2.1\times10^7$ & $1.0\times10^{12}$ &\cr
\+ \cite{churchill1975correlating} Churchill \& Chu &&\quad King && air &\hfil 0.70 &\hfil          vertical & $3.6\times10^3$ & $1.5\times10^{8}$ &\cr
\+ \cite{churchill1975correlating} Churchill \& Chu &&\quad Saunders && mercury &\hfil 0.024 &\hfil vertical & $2.7\times10^0$ & $1.7\times10^{12}$ &\cr
\+ \cite{fujii1972natural} Fujii \& Imura &&\quad $5\times10$ cm && water &\hfil 5.0 &\hfil         vertical & $7.2\times10^6$ & $1.7\times10^{8}$ &\cr
\+ \cite{fujii1972natural} Fujii \& Imura &&\quad $30\times15$ cm && water &\hfil 5.0 &\hfil        vertical & $5.2\times10^8$ & $6.1\times10^{10}$ &\cr
\+ \cite{KOBUS19953329} Hassani \& Hollands &&\quad 82~mm disk && air &\hfil 0.71 &\hfil            vertical & $1.4\times10^0$ & $2.7\times10^{5}$ &\cr
\+ \cite{KOBUS19953329} Kobus \& Wedekind &&\quad 3 sizes disk && air &\hfil 0.71 &\hfil            vertical & $6.7\times10^1$ & $1.2\times10^{4}$ &\hfil 10\% &\cr
\midrule
\+ \cite{AIHARA19722535} Aihara et al &&\quad $25\times35~\rm{cm}$ && air &\hfil 0.71 &\hfil        down & $7.2\times10^{6} $ & $ 1.0\times10^{7}$ &\cr
\+ \cite{FAW19821157} Faw \& Dullforce &&\quad 18.1~cm disk && air &\hfil 0.71 &\hfil               down & $1.1\times10^{6} $ & $ 1.6\times10^{6}$ &\hfil 1.2-2.5\% &\cr
\+ \cite{goldstein_lau_1983} Goldstein \& Lau &&\quad 2.6-20 cm square && air &\hfil 0.7 &\hfil     down & $2.5\times10^2$ & $4.5\times10^3$ &\hfil $10\%$ &\cr
\+ \cite{fujii1972natural} Fujii \& Imura &&\quad $5\times10$ cm && water &\hfil 5.0 &\hfil         down & $1.8\times10^7$ & $4.5\times10^{9}$ &\cr
\+ \cite{fujii1972natural} Fujii \& Imura &&\quad $30\times15$ cm && water &\hfil 5.0 &\hfil        down & $9.3\times10^9$ & $7.6\times10^{11}$ &\cr
\midrule
\+ \cite{fujii1972natural} Fujii \& Imura &&\quad $5\times10$ cm && water &\hfil 5.0 &\hfil         inclined &\qquad~$1.0\times10^8$ &\cr
\+ \cite{fujii1972natural} Fujii \& Imura &&\quad $30\times15$ cm && water &\hfil 5.0 &\hfil        inclined &\qquad~$1.0\times10^{10}$ &\cr
\bottomrule
}

\section{Materials and Methods}

There are robust measurements of natural convection heat and mass
transfer in the peer-reviewed literature.  {{\tabref{tab:sources}}
lists the data-sets to be compared with the present theory.  Within
each face group, all $\Ra$ have been scaled to the same
characteristic-length metric.  The ``$\pm$'' column is the
estimated convection measurement uncertainty (scatter from Lloyd
and Moran~{\cite{lloyd1974natural}}) reported in the cited source.
Both plates from Fujii and Imura~{\cite{fujii1972natural}} had
side-walls.}

  Measurements were copied from the text of Aihara
  et~al~{\cite{AIHARA19722535}}, Faw and
  Dullforce~{\cite{FAW19821157}}, and Goldstein and
  Lau~{\cite{goldstein_lau_1983}}.  The remaining heat transfer
  measurements were captured from graphs in the cited works by
  measuring the distance from each point to its graph's axes, then
  scaling to the graph's units using the ``Engauge'' software.
  Measurements obscured by other points in the graph were
  excluded.\numberedfootnote{Excluding the obscured points eliminates
    clustered points, not outliers; thus, correlations with theory
    will tend to be weaker than if all the points had been included.}

\section{Prior Work}



\subsection{Atmospheric Science}
  Renn\'o and Ingersoll~\cite{Renno.Ingersoll.JAS96} relates the
  ``convective available potential energy'' (CAPE) of a planetary
  atmosphere to heat-engine efficiency.  Atmospheric convection is
  analyzed as a four phase cyclic heat-engine.  They introduce ``total
  convective available potential energy''~(TCAPE) in terms of the
  reversible heat-engine efficiency limit $\eta=\Delta{T}/T$.
  For dry air:
$${\rm TCAPE}\approx{\eta\,c_p\,\Delta{T}}\qquad
\Delta{T}=T-T_\infty\eqdef{eqn:TCAPE}$$
  where $c_p$ is the fluid's specific heat (at constant pressure), $T$
  is the absolute temperature of the ground (heat-source), and
  $T_\infty$ is the upper atmosphere (heat-sink) absolute temperature.

  ${\rm CAPE}={\eta_N\,c_p\,\Delta{T}}$, where~$\eta_N$ is the
  heat-engine efficiency of natural convection.  Renn\'o and Ingersoll
  assert that ${\rm TCAPE}\approx2\,{\rm CAPE}$.  Thus,
  $\eta_N\approx\eta/2$.

\smallskip

The Goody~\cite{doi:10.1175/1520-0469(2003)060<2827:OTMEOD>2.0.CO;2}
cyclic heat-engine analysis splits atmospheric convection into four
phases, three of them reversible, and accounts for their energy
flows and entropy.
\smallskip
\centerline{\bf\tabdef{tab:efficiency}\quad{atmospheric convection efficiency limit}}
\moveright 0.25\hsize\vbox{\settabs 8\columns
\toprule
\+\hfil$T_\infty$ & \hfil$T$ & \hfil$\eta_A$ & \hfil$\eta_N$ & \cr
\midrule
\+\hfil240~K & \hfil300~K & \hfil10.2\% & \hfil10.0\% & \cr
\+\hfil260~K & \hfil295~K & \hfil~7.7\% & \hfil5.90\% & \cr
\+\hfil250~K & \hfil295~K & \hfil~7.7\% & \hfil7.63\% & \cr
\bottomrule
}

  \tabref{tab:efficiency} shows the sink ($T_\infty$) and source ($T$)
  temperatures, the efficiency limits from Goody ($\eta_A$),
  and~$\eta_N=\eta/2=\Delta{T}/[2\,T]$.  In the first row,
  $\eta_A$ matches $\eta_N$ within $2\%$.  In the second row, $\eta_A$
  is 30\% larger than $\eta_N$.  The third row changes $T_\infty$ from
  260~K to 250~K, causing $\eta_A$ and $\eta_N$ to match within~1\%.
  If ``260~K'' was a misprint, then both studies agree that
  $\eta_N\approx\eta/2$.

\subsection{Upward Natural Convection From a Horizontal Plate}
  For upward convection heat or mass transfer from a horizontal
  surface, prior
  works~\cite{fujii1972natural,KITAMURA2015320,lloyd1974natural,GOLDSTEIN19731025,goldstein_lau_1983}
  propose constant coefficients fitted to fractional powers of $\Ra$,
  spanning various $\Ra$ ranges.  The goal of this investigation being
  a comprehensive formula, the present theory will be compared with
  the measurements presented in these works, not with their piece-wise
  power-law approximations.

\subsection{Natural Convection From a Vertical Plate}
  The natural convection heat transfer formula developed by Churchill
  and Chu~\cite{churchill1975correlating} for
  a vertical rectangular isothermal plate of height $L$ is:
$$\Nuol^{1/2} = {0.825 + {0.387\,\Ra^{1/6}\over\bigl[1+{\left(0.492/\Pra\right)^{9/16}}\bigr]^{8/27}}}
 \qquad 1\le \Ra\le 10^{12}\eqdef{eqn:Churchill-Chu}$$

\subsection{Downward Natural Convection From a Horizontal Plate}
  Schulenberg~\cite{SCHULENBERG1985467} derives a formula for
  convection below a level isothermal strip of width $2\,L$.
  Proposed is a corrected\numberedfootnote{The
  Schulenberg~\cite{SCHULENBERG1985467} heat transfer formula for
  an isothermal strip was:
$$\Nuol={0.571\,\Ra^{1/5}\,\Pra^{1/5}\over\bigl[1+1.156\,\Pra^{3/5}\bigr]^{1/3}}
 ={0.544\,\Ra^{1/5}\over\bigl[1+(0.785/\Pra)^{3/5}\bigr]^{1/3}}\eqdef{eqn:Schulenberg-strip-orig}$$
  1.156 is the only 4-digit coefficient in the paper's isothermal
  plate correlations; the others have 3 significant digits.
  \figref{fig:Ra-factor} of the present work compares the effective~$\Ra$
  scale factor of all four of these formulas; the
  ``1.156~Schulenberg strip~$1/\Xi_?$''
  trace is 40\% lower than the others at $\Pra\ll1$.}
  formula and its equivalent, normalized so that $\Pra$ appears only
  in the denominator:
$$\Nuol={0.631\,\Ra^{1/5}\,\Pra^{1/5}
  \over\bigl[1+1.56\,\Pra^{3/5}\bigr]^{1/3}}
  ={0.544\,\Ra^{1/5}\over\bigl[1+\left(0.477/\Pra\right)^{3/5}\bigr]^{1/3}}
  \eqdef{eqn:Schulenberg-strip}$$

  Schulenberg also gives a formula for downward convection from a
  level isothermal disk using its radius as the characteristic-length.
  The expression on the right side is its equivalent normalized form:
$$\Nuol_r={0.705\,\Ra^{1/5}\,\Pra^{1/5}
  \over\bigl[1+1.48\,\Pra^{3/5}\bigr]^{1/3}}
  ={0.619\,\Ra^{1/5}\over\bigl[1+\left(0.520/\Pra\right)^{3/5}\bigr]^{1/3}}
  \eqdef{eqn:Schulenberg-disk}$$


\section{Unenclosed Heat-Engine}

  Although most textbook heat-engine analyses are of cyclic
  heat-engines, a continuous process can also convert a temperature
  difference into mechanical work, which qualifies it as a
  heat-engine.

  Consider a large vertical column of still, dry air having molar mass
  $M$ under the influence of gravitational acceleration $g$.
  Initially, the air will be in equilibrium, with uniform absolute
  temperature $T_\infty$ and a pressure profile~$P$ which decays
  exponentially with altitude~$z$:
$$P(z)=P_0\,\exp\left({-z\,g\,M\over\overline{R}\,T_\infty}\right)\eqdef{eqn:P(z)}$$

  Where $\overline{R}$ is the universal gas constant, the ideal gas
  law finds the density~$\rho$ of a parcel of air:
$$\rho={M\,P\over\overline{R}\,T}\eqdef{eqn:gas-law}$$

  A heated parcel of volume $V$ has density
  $\rho_h=M\,P/\left[[T_\infty+\Delta{T}]\,\overline{R}\right]$.  The
  buoyancy force on it is:
$$[\rho-\rho_h]\,g\,V={g\,V\,M\,P\,\Delta{T}\over\overline{R}\,T_\infty\,[T_\infty+\Delta{T}]}\eqdef{eqn:force=}$$

  Where $0<\Delta{T}\ll T_\infty$, and
  $c_p$ is the specific heat (at constant pressure) of the
  fluid, $\Delta{Q}=c_p\,\rho\,V\,\Delta{T}$ is the heat
  required to raise the parcel temperature from $T_\infty$ to
  $T=T_\infty+\Delta{T}$.
  The force on the parcel is:
$$[\rho-\rho_h]\,g\,V=\left[{M\,P\over\rho\,\overline{R}\,T}\right]\,{g\,\Delta{Q}\over c_p\,T}
  = {g\,\Delta{Q}\over c_p\,T}={g\,\rho\,V\,\Delta{T} \over T}\eqdef{eqn:force}$$

  As it rises, the parcel's state changes.  Temperature, volume, and
  pressure are three variables having two degrees of freedom from
  formula~\eqref{eqn:gas-law}.  For a large vertical column of still, dry
  air,
  Fermi~\cite{fermi1956thermodynamics} teaches:

{\narrower
  ``Since air is a poor conductor of heat, very little heat is
  transferred to or from the expanding air, so that we may consider
  the expansion as taking place adiabatically.''
  \par}

  Hence, the temperature of a parcel of dry air drops
  $g/c_p\approx9.8\rm~K$ per kilometer of altitude gain.
  In a column having initially uniform temperature, a heated parcel
  will rise until its temperature drops to $T_\infty$.

  From the conservation of mass, $\rho\,V=\rho_0\,V_0$, where $\rho_0$
  and $V_0$ are the density and volume at altitude $z=0$.  The
  maximum work $W$ which can be extracted from a buoyant parcel is
  the integral of upward force formula~\eqref{eqn:force} with respect
  to altitude~$z$ above the heated plate:
$$W=\int_0^{\Delta{T}\,c_p/g}{[\Delta{T}-z\,g/c_p]\,g\,\rho_0\,V_0 \over T}\,\diff{z}
   ={g\,\Delta{Q}\over c_p\,T}\,{\Delta{T}\,c_p\over2\,g}
   ={\Delta{Q}\,\Delta{T}\over2\,T}\eqdef{eqn:W=}$$

  The thermodynamic efficiency ($W/\Delta{Q}$) of this ideal
  convection heat-engine will be the thermodynamic efficiency limit
  for external convection, $\eta_N$:
$$\eta_N={W \over \Delta{Q}}={\Delta{T}\over2\,T}\eqdef{eqn:eta=}$$
  Note that $\eta_N$ is $1/2$ of the (Carnot) reversible heat-engine
  efficiency limit~$\eta=\Delta{T}/T$.

  This derivation was for adiabatic gases whose coefficient of
  thermal expansion~$\beta=1/T$.  More generally:
$$\eta_N={\beta\,{\Delta{T}}\over2}\eqdef{eqn:beta}$$

  The system being in continuous operation, instead of energies $W$
  and $\Delta{Q}$, power fluxes (${\rm W/m^2}$) are of interest.  The
  powers per heated plate area are $I_k$ for the kinetic flux of the
  fluid and $I_p$ for the plate total, which is also the convective
  power flux.  The thermodynamic efficiency of a steady-state
  convection process is ${I_k/I_p}$, which the second law of
  thermodynamics constrains so that:
$${I_k\over I_p}\le\eta_N\eqdef{eqn:eta}$$ 

\section{Dimensional Analysis}


  Additional fluid properties used in this investigation are thermal
  conductivity~$k$, kinematic viscosity~$\nu$, and thermal
  diffusivity~$\alpha=k/[\rho\,c_p]$.  $\overline{h}$~is the average
  convective surface conductance, with units ${\rm W/(m^2\cdot K)}$.

  ``Scalable'' heat transfer equations relate named, dimensionless
  ``variable groups'', which themselves are functions of variables and
  other variable groups.
  ``Dimensional analysis'' discovers these dimensionless variable
  groups and their scalable relationships.


  Nusselt's dimensional analysis of natural convection (from Lienhard
  and Lienhard~\cite{ahtt5e}) employs four variable groups: average
  Nusselt number $\Nuol$, Prandtl number $\Pra$, $\Pi_3$, and $\Pi_4$.
$$\Nuol\equiv{\overline{h}\,L\over k}={I_p\,L\over\Delta{T}\,k},
  \qquad\Pra\equiv{\nu\over\alpha},
  \qquad\Pi_3\equiv{L^3\over\nu^2}\,g={L\,g\over[\nu/L]^2},
  \qquad\Pi_4\equiv\beta\,\Delta{T}\eqdef{eqn:variable-groups}$$
The $\Nuol={I_p\,L/[\Delta{T}\,k]}$ equivalence was added for this
investigation.

  From these variable groups come the dimensionless Grashof number
  ($\Gr$) and Rayleigh number ($\Ra$):
$$\Gr\equiv\Pi_3\,\Pi_4={\beta\,\Delta{T}\,g\,L^3\over\nu^2},\qquad
  \Ra\equiv\Gr\,\Pra={\beta\,\Delta{T}\,g\,L^3\over\alpha\,\nu}\eqdef{eqn:Gr}$$

  From formula~\eqref{eqn:beta} and $\Pi_4\equiv\beta\,\Delta{T}$ from
  formula~\eqref{eqn:variable-groups}:
$$\eta_N={\Pi_4\over2}\eqdef{eqn:eta-pi}$$

  Let~$\Pi_5$ be the heat transport capacity per kinetic energy ratio.
  $\Pi_5$~increases with~$\beta$, $g$, $L$, and~$\Pra$:
$$\Pi_5={\beta\,g\,L\,\Pra^2/c_p}\eqdef{eqn:Pi_5}$$
  The denominator $c_p$, canceling one of the two factors of $c_p$ in
  $\Pra^2$, makes~$\Pi_5$ dimensionless.

  Two variable groups with power flux units (${\rm W/m^2}$) will prove
  useful:
$$\Phi_p={k\,\Delta{T}\over L}
  \qquad\Phi_k=\left[{\nu\over L}\right]^3\rho\,{\Pi_4\,\Pi_5}\eqdef{eqn:flux-groups}$$

\section{Combining Heat Transfers}

  Conduction and convection are both heat transfer processes.

  There is an unnamed form which appears frequently in heat transfer
  formulas:
$$F^p(\xi)=F_0^p(\xi)+F_\infty^p(\xi)\eqdef{eqn:mixing}$$

  Churchill and Usagi~\cite{AIC:AIC690180606} wrote that such formulas
  are ``remarkably successful in correlating rates of transfer for
  processes which vary uniformly between these limiting cases.''

  A value of $p>1$ models competitive processes; the combined transfer
  rate is between the larger constituent rate and their sum.

  $p=1$ models independent processes; the combined transfer rate is
  the sum of the constituent rates.

  The Churchill and Chu formula~\eqref{eqn:Churchill-Chu} for vertical
  plates has the form of equation~\eqref{eqn:mixing} with~$p=1/2$.
  Natural convection requires some conduction to heat the fluid.  This
  is consistent with cooperating processes having~$0<p<1$; when both
  are transferring, the combination is larger than their sum.

  With $F_0(\xi)\ge0$ and $F_\infty(\xi)\ge0$, taking the~$p$th root
  of both sides of equation~\eqref{eqn:mixing} yields a vector-space
  functional form known as the $\ell^p$-norm, which is notated
  $\|F_0~,~F_\infty\|_p$:
$$\left\|F_0~,~F_\infty\right\|_p=\left(|F_0|^p+|F_\infty|^p\right)^{1/p}\eqdef{eqn:l^p}$$


\section{Upward-Facing Circular Plate}

\subsection{Characteristic-Length}
  For a horizontal upward-facing plate, \figref{fig:above-flow}
  shows that natural convection pulls fluid from the edges into a
  central plume.  The characteristic-length should be a function of
  radial distance from the edges to the center.  This is
  accomplished by using the area-to-perimeter ratio~$\Ls$ as the
  characteristic-length~$L$.

  Lloyd and Moran~\cite{lloyd1974natural} measured upward convection
  from horizontal disks, rectangles, and right triangles
  having aspect ratios between~1:1 and~10:1.  They wrote:
        ``It is immediately obvious that within the scatter of the data,
        approximately $\pm5$ percent, the data from all plan-forms are
        correlated through the use of $\Ls$,~\dots''

  Goldstein et al~\cite{GOLDSTEIN19731025} found that $\Ls$ correlated
  their measurements with aspect ratios between~1:1 and~7:1.

\subsection{Conduction}
  Consider a horizontal disk with its upper face, having area $A$,
  heated to $T_\infty+\Delta{T}$.  Its $L=\Ls$ is $1/2$ of its radius.
  The power flowing from an object into a stationary, uniform medium
  is $q=S\,k\,\Delta{T}$, where $k$ is thermal conductivity and $S$ is
  the conduction shape factor (having length unit).
  For one side of a disk, Incropera, DeWitt, Bergman, and
  Lavine~\cite{bergman2007fundamentals} gives $S=2\,D~(=8\,\Ls)$.
  Converting conduction power flux $I_p=q/A$ into conduction Nusselt
  number $\Nuzs=\Nuol$ from formula~\eqref{eqn:variable-groups}:
$$I_p\,A=\Nuzs\,A\,{k\,\Delta{T}\over \Ls}=q=S\,k\,\Delta{T}
  \qquad\Nuzs={S\,\Ls\over A}={8\,\Ls\,\Ls\over\pi\,[2\,\Ls]^2}
  ={2\over\pi}\approx0.637\eqdef{eqn:Nu0*}$$

\subsection{Kinetic Flux}
  Fluid heated near the plate converts thermal energy into kinetic
  energy by accelerating upward.  Fluid accelerating upward spreads
  apart, pulling fluid horizontally to maintain its density.
  At some elevation~$z_t$, the fluid no longer accelerates upward
  (otherwise, its velocity would be unbounded) and the horizontal flow
  is negligible, which is marked by the dashed line
  in \figref{fig:above-flow}.

  An ideal turbine at elevation $z_t$ would capture the upward kinetic
  energy of the plume.
  The kinetic power through the aperture would be ${\rho\,A\,u\,u^2/2}$,
  where~$u$ is the plume upward velocity; its flux, ${\rho\,u^3/2}$.

  Vertical acceleration pulls fluid horizontally at elevations between 0
  and $z_t$.  ``Fig.~14(f)'' of Fujii and Imura~\cite{fujii1972natural}
  shows that horizontal velocities are fairly uniform within that
  span.  The kinetic flux should be proportional to ${\rho\,u^3}/2$
  scaled by~$z_t$.  $z_t$ grows with~$u$, but shrinks with kinematic
  viscosity~$\nu$ because of viscous losses.  $u/\nu$ has reciprocal
  length units, while ${\rho\,u^3/2}$ already has power flux units.
  This suggests scaling ${\rho\,u^3/2}$ by a (dimensionless) Reynolds
  number $\Rey=u\,L/\nu$, which is used extensively for modeling
  forced flows.  Let $\Rey_i=u\,L_i/\nu$, where $L_i$ is the average
  length of flow parallel to the plate.  For the upward-facing plate,
  $L_i=2\,L=2\,\Ls$; hence $\Rey_i=2\,\Rey$, leading to a kinetic
  power flux $\Rey\,{\rho\,u^3}$.
  From the $\Pi_5$ dimensional analysis,
  $\Rey_i\,\Pi_5\,[{\rho\,u^3}/2]$ is the maximum heat flux which
  could be transported by the flow induced by~$u$.  Multiplying this
  heat flux by $\Pi_4/2$ yields the maximum kinetic flux $I_k$ which
  could result from natural convection:
$$I_k={\Rey_i}\,{\rho\,u^3\over2}\,{\Pi_4\over2}\,\Pi_5
     ={\rho\,L\over2\,\nu}\,{u^4}\,{\Pi_4\,\Pi_5}\qquad
 u=\left[{2\,\nu\,I_k\over\rho\,L\,\Pi_4\,\Pi_5}\right]^{1/4}\eqdef{eqn:I_k-up}$$

\subsection{Plate Flux}
  With upward convection pulling fluid horizontally from the disk's
  perimeter, heat transfer near the perimeter is more
  flow-induced than it is conduction.  If the flow were parallel, 1/2
  of the plate area would be considered flow-induced.  If the flow
  were radial, $1/4$ would be considered flow-induced.  However, the
  square plate photographs in Kitamura et al~\cite{KITAMURA2015320}
  show plumes as a network of connected ridge segments, not a central
  cone.  An intermediate allocation is needed.  The geometric mean of
  $1/2$ and ${1/4}$ is $\sqrt{1/8}$.  Hence, $\sqrt{1/8}\approx0.354$
  of the plate is designated as flow-induced,
  $[1-\sqrt{1/8}]\approx0.646$ of the plate as conduction.

  Heat transfer from the flow-induced part of the plate will be
  proportional to~$\Nuzs$, $\Ls$, and formula~\eqref{eqn:I_k-up}~$u$ in
  the dimensionless expression
  $\Nuzs\,{u\,\Ls/[\sqrt{8}\,\nu]}=\Nuzs\,\Rey/\sqrt{8}$.  As
  cooperating processes, conductive and flow-induced heat transfers
  combine using the $\ell^{1/2}$-norm.  Solving for plate power
  flux~$I_p$ from formula~\eqref{eqn:variable-groups}:
$$I_p={k\,\Delta{T}\over L}\,\Nuzs\,\left\|1-{1\over\sqrt{8}}~,~{\Rey\over\sqrt{8}}\right\|_{1/2}
     ={k\,\Delta{T}\over L}\,\Nuzs\,\left\|1-{1\over\sqrt{8}}~,~{L\over\sqrt{8}\,\nu}\,\left[{2\,\nu\,I_k\over\rho\,L\,\Pi_4\,\Pi_5}\right]^{1/4}\right\|_{1/2}
     \eqdef{eqn:I_p-up}$$



  Assume $\Rey\gg\sqrt{8}$, so that the $1-\sqrt{1/8}$ term can be ignored.
  From definitions~\eqref{eqn:flux-groups} collect~$\Phi_k$ and~$\Phi_p$ terms:
$$I_p=\Phi_p\,{\Nuzs\over\sqrt{8}}\,\left[{2\,I_k\over\Phi_k}\right]^{1/4}\eqdef{eqn:I_p-up3}$$

  The $I_k$ upper-bound can be found by combining~$I_p$
  formula~\eqref{eqn:I_p-up3} with~$\eta_N$ formulas~\eqref{eqn:eta}
  and~\eqref{eqn:eta-pi}:
$$I_k\le{\Pi_4\over2}\,I_p
  ={\Phi_p\,\Pi_4\,\Nuzs\over\sqrt{8}}\,\left[{I_k\over8\,\Phi_k}\right]^{1/4}\eqdef{eqn:I_k<'}$$

  Dividing both sides of formula~\eqref{eqn:I_k<'} by $I_k^{1/4}$, then
  raising both sides to the 4/3 power, isolates $I_k$:
$$I_k\le\left[{\Phi_p\,\Pi_4\,\Nuzs\over\sqrt{8}}\right]^{4/3}\left[{1\over8\,\Phi_k}\right]^{1/3}
          =\Phi_p\,\left[{\Nuzs\over\sqrt{8}}\right]^{4/3}\left[{\Phi_p\,\Pi_4\over\Phi_k}\right]^{1/3}\,{\Pi_4\over2}
\eqdef{eqn:I_k<}$$



  In the absence of obstruction,
  $I_k$ and $I_p$ will increase to the maximum allowed by upper-bound
  formula~\eqref{eqn:I_k<}.  Substituting $I_k$ from formula~\eqref{eqn:I_k<}
  into formula~\eqref{eqn:I_p-up3}
  yields the asymptotic formula for~$I_p$:
$$I_p=\Phi_p\,\left[{\Nuzs\over\sqrt{8}}\right]^{4/3}\left[{\Phi_p\,\Pi_4\over\Phi_k}\right]^{1/3} \eqdef{eqn:I_p=*}$$

  Both $I_p$ and $I_k$ have $\root3\of{\Phi_p\,\Pi_4/\Phi_k}$ factors.
  How does ${\Phi_p\,\Pi_4/\Phi_k}$ relate to formula~\eqref{eqn:Gr}~$\Ra$?
$$\eqalignno{{\Phi_p\,\Pi_4\over\Phi_k}={k\,\Delta{T}\over\rho\,L}\,{L^3\over\nu^3}\,{\beta\,g\,L\over c_p}\,{\nu^2\over\alpha^2}
  &={\beta\,\Delta{T}\,g\,L^3\over\alpha\nu}=\Ra&\eqdef{eqn:Ra}\cr
  I_p=\Phi_p\left[{\Nuzs\over\sqrt{8}}\right]^{4/3}\Ra^{1/3}=\Phi_p\,\Nuols\qquad
  \Nuols&={\Nuzs^{4/3}\over4}\Ra^{1/3}
  \approx0.137\,\Ra^{1/3}&\eqdef{eqn:Nu*=}\cr}$$



  Restoring the $\ell^{1/2}$-norm from equation~\eqref{eqn:I_p-up} into
  equation~\eqref{eqn:Nu*=} yields the comprehensive formula for natural
  convection heat transfer from an external, horizontal plate's
  isothermal upper face:
$$\Nuols=\left\|\Nuzs\,\left[1-{1\over\sqrt{8}}\right]~,~{\Nuzs^{4/3}\over4}\,\Ra^{1/3}\right\|_{1/2}
 \approx\left[0.642+0.370\,\Ra^{1/6}\right]^2\eqdef{eqn:upward}$$

\subsection{Measurement}
 $\Nuols$ formula~\eqref{eqn:upward} assumes unobstructed flow.
 A completely unobstructed apparatus is difficult to build.
 Measurements smaller than $\Nuols$ are expected.

 Measurement bias and uncertainty can result in values slightly larger
 than~$\Nuols$.

\section{Upward-Facing Measurements}

\vbox{\settabs 1\columns
\+\hfil\figscale{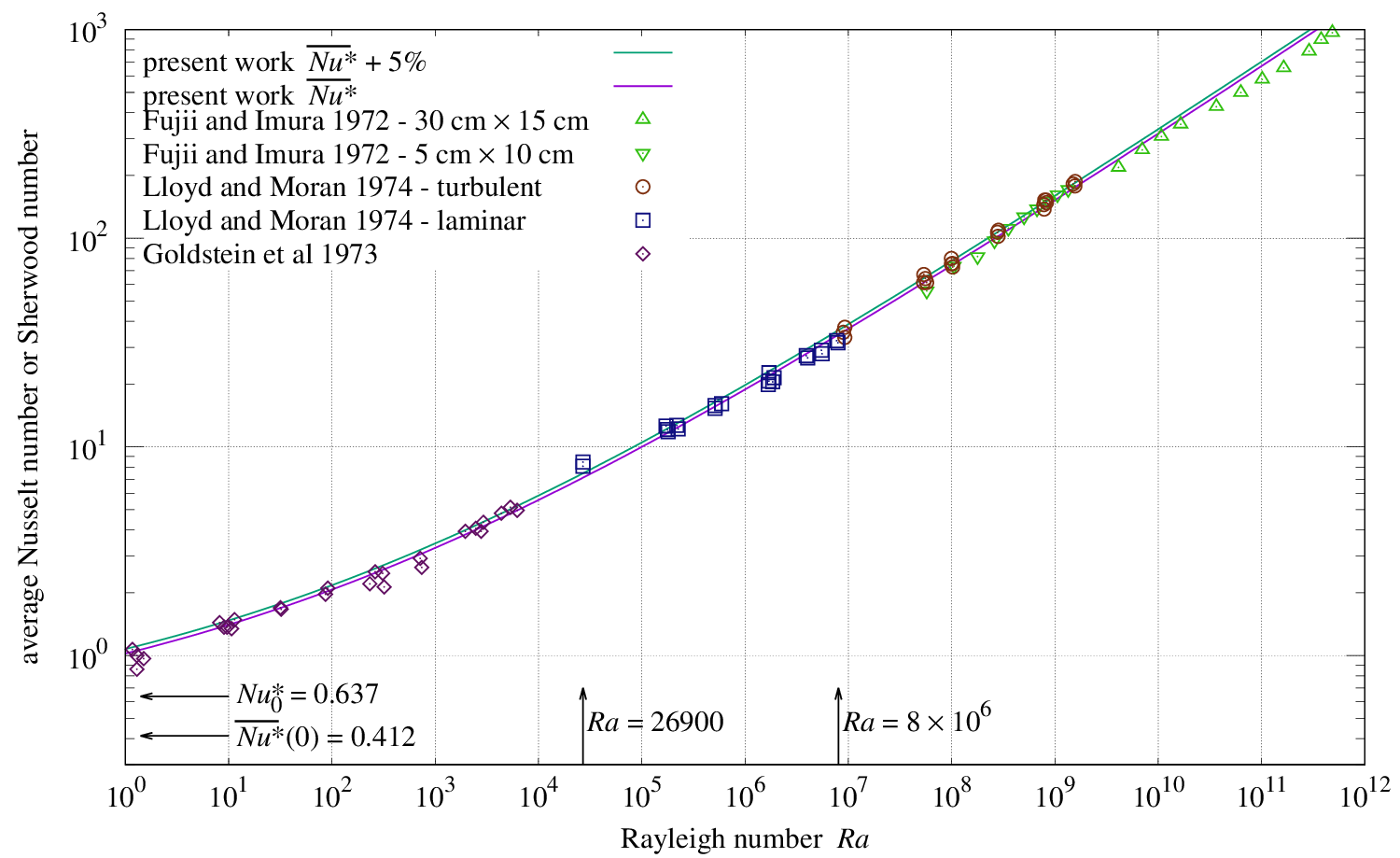}{468pt}&\cr
\+\hfil{\bf\figdef{fig:nuhup}\quad{upward convection heat transfer from horizontal plate}}&\cr
}
\medskip
\centerline{\bf\tabdef{tab:up-conformance}\quad{upward convection heat transfer from horizontal plate}}
\vbox{\offinterlineskip \settabs 9\columns
\toprule
\+\hfil source&&data-set&\hfil$\Pra|Sc$ &\hfil face &\quad~formula&\hfil RMSRE &\quad bias\qquad scatter&\qquad\quad count&\cr
\midrule
\+ Fujii and Imura~\cite{fujii1972natural} -- $30~\rm{cm}\times15~\rm{cm}$&&&\hfil 5.0 &\hfil   up &\quad\eqref{eqn:upward}~$\Nuols$&\hfill 12.0\%~\quad&\hfill$-11.4\%$\qquad& 4.0\%\hfill11&\cr
\+ Fujii and Imura~\cite{fujii1972natural} -- $5~\rm{cm}\times10~\rm{cm}$&&& \hfil 5.0 &\hfil   up &\quad\eqref{eqn:upward}~$\Nuols$&\hfill  5.0\%~\quad&\hfill$-0.4\%$\qquad& 5.0\%\hfill10&\cr
\+ Lloyd and Moran~\cite{lloyd1974natural} -- electrochemical&&&\hfil 2200 &\hfil               up &\quad\eqref{eqn:upward}~$\Nuols$&\hfill  4.9\%~\quad&\hfill$+0.7\%$\qquad& 4.8\%\hfill39&\cr
\+ Goldstein et al~\cite{GOLDSTEIN19731025} -- sublimation&&&\hfil 2.50 &\hfil                  up &\quad\eqref{eqn:upward}~$\Nuols$&\hfill  7.2\%~\quad&\hfill$-2.3\%$\qquad& 6.8\%\hfill26&\cr
\bottomrule
}
\medskip


Lloyd and Moran~\cite{lloyd1974natural} estimated 5\% scatter for
their data.  Two ``Lloyd and Moran 1974 -- laminar'' points at
$\Ra\approx26900$ have values 14\% and 19\% larger than~$\Nuols$
in \figref{fig:nuhup}.

$\Ra\approx26900$ was the smallest~$\Ra$ measured by Lloyd and Moran;
the next smallest $\Ra=171750$ was~6.4 times larger.  Range extremes
are often the most susceptible to measurement bias.  Having excesses
several times larger than~5\%, the points at $\Ra\approx26900$ should
be excluded as outliers.

Excluding the two largest and two smallest measurements relative
to~$\Nuols$, the Lloyd and Moran measurements have a 4.9\%
root-mean-squared relative error (RMSRE) from~$\Nuols$.  This is a
close match spanning four orders of magnitude of $\Ra$ which includes
the laminar-turbulent transition at $\Ra\approx8\times10^6$.

\subsection{RMSRE}
 RMSRE gauges the fit of measurements $g(\Ra)$ to formula $f(\Ra)$,
 giving each measurement equal weight.  The root-mean-squared error
 of~$g(\Ra)$ relative to~$f(\Ra)$ at~$n$ points~$\Ra_j$ is:
$$\sqrt{{1\over n}\sum_{j=1}^n\left|{g(\Ra_j)\over f(\Ra_j)}-1\right|^2}\eqdef{eqn:RMSRE}$$

 \tabref{tab:up-conformance} also splits RMSRE into bias and scatter.  The
 root-sum-squared of bias and scatter is RMSRE.


\medskip

 Note that $\Nuols(0)<\Nuzs$ in \figref{fig:nuhup}.  $\Nuols$~models
 convection; it does not extend to static conduction.

 The Fujii and Imura $30~\rm{cm}\times15~\rm{cm}$ upward-facing
 data-set is revisited
 in \ref{Upward-Facing Rectangular Plate With Side-Walls}.

\section{Vertical Rectangular Plate}

\subsection{Characteristic-Length and Conduction}
  The vertical characteristic-length $L'=L$ is the plate's height.
  Conduction constant $\Nuzq$ will not depend on the plate's width.
  The asymptotic case is a strip, an infinitely wide rectangle.

  Conduction shape factors are not well-defined with unbounded source
  areas, but Nusselt numbers can be.  Fortunately, $\Nuzq$ for a strip
  can be related to square plate~$\Nuz$.
  Incropera
  et al~\cite{bergman2007fundamentals} gives a dimensionless shape
  factor $q_{SS}^*=0.932$ for both faces of a rectangular plate.  For
  one face of an $L\times L$ square plate:
$$\eqalignno{A_0=2\,A=2\,L^2
 \qquad L_0=\sqrt{A_0\over4\,\pi}={L\over\sqrt{2\,\pi}}
 \qquad &q={q_{SS}^*\over2}\,k\,\Delta{T}\,{A\over L_0}={q_{SS}^*}\,k\,\Delta{T}\,L\,{\sqrt{\pi\over2}}&\eqdef{eqn:q=}\cr
 \Nuz\,A\,{k\,\Delta{T}\over L}={q}={q_{SS}^*}\,k\,\Delta{T}\,L\,{\sqrt{\pi\over2}}
  &\qquad\Nuz={q_{SS}^*}\,{\sqrt{\pi\over2}}\approx1.168&\eqdef{eqn:Nu_0}\cr}$$

  Strip conduction $\Nuzq$ must distribute over one dimension
  (vertical) what square plate conduction $\Nuz$ distributes over two:
$$\Nuzq={\Nuz^2}={q_{SS}^*}^2\,{\pi\over2}\approx1.363\eqdef{eqn:Nu0'}$$

\subsection{Kinetic and Plate Fluxes}
  Fluid is pulled horizontally before rising into a plume at
  the plate.  The upward flow is parallel; plate area is treated as
  1/2~flow-induced, 1/2~conduction.  The average length of contact
  with the plate is $L/2$, resulting in the~$I_k$ factor
  $\Rey/2={u\,L/[2\,\nu]}$.  Fluid heated by the plate accelerates
  upward along its surface.  This reduces the effective length of
  contact by 1/2, resulting in $\Rey/4$ as the heat transfer factor
  in formula~\eqref{eqn:I_p'}.  The kinetic and plate power fluxes are:
$$\eqalignno{I_k={\Rey\over2}\,{\rho\,u^3\over2}\,{\Pi_4\over2}\,\Pi_5
     ={L\over\nu}\,{\rho\,u^4\over8}\,{\Pi_4\,\Pi_5}\qquad
 &u=\left[{\nu\over L}\,{8\,I_k\over\rho\,\Pi_4\,\Pi_5}\right]^{1/4}&\eqdef{eqn:I_k'}\cr
  I_p={k\,\Delta{T}\over L}\,\Nuzq\,\left\|{1\over2}~,~{1\over2}\,{\Rey\over4}\right\|_{1/2}
     ={k\,\Delta{T}\over L}\,\Nuzq\,&\left\|{1\over2}~,~{L\over8\,\nu}\,\left[{\nu\over L}\,{8\,I_k\over\rho\,\Pi_4\,\Pi_5}\right]^{1/4}\right\|_{1/2}
     &\eqdef{eqn:I_p'}\cr}$$

  Assume $\Rey={u\,L/\nu}\gg1$ and ignore the conduction term; collect
  $\Phi_p$ and $\Phi_k$ terms from definitions~\eqref{eqn:flux-groups}:
$$I_p={\Phi_p\over8}\,\Nuzq\,\left[{8\,I_k\over\Phi_k}\right]^{1/4}\eqdef{eqn:I_p-v}$$

  The $I_k$ upper-bound can be found by combining~$I_p$
  formula~\eqref{eqn:I_p-v} with~$\eta_N$ formulas~\eqref{eqn:eta}
  and~\eqref{eqn:eta-pi}:
$$I_k\le{\Pi_4\over2}\,I_p
  ={{\Phi_p}\,\Pi_4\,\Nuzq\over16}\,\left[{8\,I_k\over\Phi_k}\right]^{1/4}
  \eqdef{eqn:I_k'<}$$

  Dividing both sides of formula~\eqref{eqn:I_k'<} by $I_k^{1/4}$, then
  raising both sides to the 4/3 power, isolates $I_k$:
$$I_k\le\left[{\Phi_p\,\Pi_4\,\Nuzq\over16}\right]^{4/3}\left[{8\over\Phi_k}\right]^{1/3}
     =\Phi_p\,{\Nuzq^{4/3}\over8\,\root3\of2}\,\left[{\Phi_p\,\Pi_4\over\Phi_k}\right]^{1/3}\,{\Pi_4\over2}
\eqdef{eqn:I_k<v}$$

  The plate partially obstructs flow; $\Nuolq$ will be an upper-bound.
  Reduce to $\Ra$ using equation~\eqref{eqn:Ra}:
$$\eqalignno{I_p\le{2\over\Pi_4}\,I_k
    =\Phi_p\,{\Nuzq^{4/3}\over8\,\root3\of2}\,\left[{\Phi_p\,\Pi_4\over\Phi_k}\right]^{1/3}
    &=\Phi_p\,{\Nuzq^{4/3}\over8\,\root3\of2}\Ra^{1/3}\ge\Phi_p\,\Nuolq&\eqdef{eqn:I_p'=}\cr
  \Nuolq\le{\Nuzq^{4/3}\over8\,\root3\of2}\Ra^{1/3}&\approx0.150\,\Ra^{1/3}&\eqdef{eqn:Nu'<}}$$

  Reintroduce the $\ell^{1/2}$-norm into formula~\eqref{eqn:Nu'<}:
$$\Nuolq\le\left\|{\Nuzq\over2}~,~{{\Nuzq}^{4/3}\over8\,\root3\of2}\Ra^{1/3}\right\|_{1/2}
 \approx\left[0.826+0.387\,\Ra^{1/6}\right]^2\eqdef{eqn:vertical<}$$

  For large $\Pra$, the Churchill and Chu equation~\eqref{eqn:Churchill-Chu}
  reduces to $\Nuol=\bigl[0.825+0.387\,\Ra^{1/6}\bigr]^2$.

  The $\Ra$ term's denominator in equation~\eqref{eqn:Churchill-Chu} is
  always greater than 1; it can only reduce the magnitude of~$\Nuol$.
  Therefore, formula~\eqref{eqn:Churchill-Chu} satisfies $\Nuolq$
  upper-bound formula~\eqref{eqn:vertical<}.

\section{Downward-Facing Rectangular Plate}

  \figref{fig:below-flow} shows flow in two plumes on horizontally
  opposite sides of the plate; downward-facing
  characteristic-length~$L_R$ is $1/2$ of the shorter plate edge.
  Compared with the vertical convection strip, $L_R=L'/2$ and
  $\Nuz=\Nuzq/2$.  Plate areas are treated as 1/2~flow-induced,
  1/2~conduction.

  For upward-facing and vertical plates, the induced flow brings
  unheated fluid into contact with the plate, which is responsible for
  amplifying convection.
  In contrast, fluid below the downward-facing plate is warmed by
  conduction through the fluid above it.  The outward creep
  immediately below the plate stays in contact until it reaches a
  plate edge.  The fluid's temperature profile differs little from
  static conduction.  Thus, static and dynamic heat transfers are
  combined using the $\ell^1$-norm.

  \figref{fig:below-flow} shows convective flow experiencing three
  $90^\circ$ changes of direction.  Two horizontal accelerations and
  decelerations of flow introduce two factors of $2\,\Rey=2\,u\,L/\nu$
  into~$I_k$.  The short upward acceleration and deceleration of flow
  below the plate is the only such occurrence among the three plate
  orientations.  It slightly opposes buoyant flow because fluid
  immediately below the plate is less dense than fluid moving upward
  to replace it.  There being no appropriate vertical distance,
  $\Rey$~is used for the third factor in~$I_k$.
$$I_k=4\,\Rey^3\,{\rho\over2}\,{u^3}\,{\Pi_4\over2}\,\Pi_5
     ={L^3\over\nu^3}\,{2\,\rho\,u^6}\,{\Pi_4\,\Pi_5}\qquad
 u=\left[{\nu^3\over L^3}\,{I_k\over2\,\rho\,\Pi_4\,\Pi_5}\right]^{1/6}\eqdef{eqn:I_k=R}$$

  All of the lower face is in contact with horizontal flow; the heat
  transfer factor is~$2\,\Rey$:
$$I_p={k\,\Delta{T}\over L}\,{\Nuz}\,\left[{1\over2}+{2\,\Rey\over2}\right]
     ={k\,\Delta{T}\over L}\,{\Nuz}\,\left[{1\over2}+{L\over\nu}\,\left[{\nu^3\over L^3}\,{I_k\over2\,\rho\,\Pi_4\,\Pi_5}\right]^{1/6}\right]
     \eqdef{eqn:I_p=R}$$

  Assume ${u\,L/\nu}\gg1$ and ignore the conduction term; collect
  $\Phi_p$ and $\Phi_k$ terms; then solve for~$I_p$:
$$I_p={\Phi_p}\,{\Nuz}\,\left[{I_p\over2\,\Phi_k}\right]^{1/6}
 \qquad I_p={\Phi_p}\,{\Nuz^{6/5}\over2^{1/5}}\,\Ra^{1/5}=\Phi_p\,\Nuol_R
 \eqdef{eqn:Nu0R}$$

  The plate obstructs heated flow from rising; thus, $\Nuol_R$ will be
  an upper-bound.  Solving equation~\eqref{eqn:Nu0R} for~$\Nuol_R$,
  restoring the $\ell^1$-norm (addition), and substituting~$\Nuzq/2$
  for~$\Nuz$:
$$\Nuol_R\le{\Nuzq\over4}+{\Nuzq^{6/5}\over2^{7/5}}\,\Ra^{1/5}\approx0.341+0.550\,\Ra^{1/5}
 \eqdef{eqn:down<}$$

  For large $\Pra$, the Schulenberg strip convection
  formula~\eqref{eqn:Schulenberg-strip} reduces to ${0.544\,\Ra^{1/5}}$.
  The formula~\eqref{eqn:down<} coefficient 0.550 is 1.1\% larger
  than 0.544.  Being greater than 1, the denominator of
  formula~\eqref{eqn:Schulenberg-strip} can only reduce the magnitude of
  $\Nuol$.  Therefore, formula~\eqref{eqn:Schulenberg-strip} satisfies
  $\Nuol_R$ upper-bound formula~\eqref{eqn:down<}.


\section{Generalization}

  These three derivations can be generalized to
  equation~\eqref{eqn:general} via formulas~\eqref{eqn:I_p-and-I_k}.
  $L$ is characteristic-length; $\Nuz$~is static conduction; $E$~is
  the count of $90^\circ$ changes in direction of fluid flow; $B$~is
  the sum of the mean lengths of flows parallel to the plate divided
  by~$L$; $C$~is the plate area fraction responsible for flow induced
  heat transfer; $D$~is the effective length of heat transfer contact
  with the plate divided by~$L$; $p=1/2$ when $E=1$ without
  side-walls; otherwise $p=1$.
$$\eqalignno{I_k=&\,{B\,\Rey^E}\,{\rho\,u^3\over2}\,{\Pi_4\over2}\,\Pi_5 \qquad
  I_p={k\,\Delta{T}\over L}\,\Nuz\,\left\|1-C, {C\,D\,\Rey}\right\|_p&\eqdef{eqn:I_p-and-I_k}\cr
  &\Nuol=\left\|\Nuz\,\left[1-C\right], \root{2+E}\of{\left[{C\,D\,\Nuz}\right]^{3+E}{{2\over B}\,\Ra}}~\right\|_p&\eqdef{eqn:general}\cr}$$

\centerline{\bf\tabdef{tab:derivation parameters}\quad{derivation parameters}}
\moveright 0.091\hsize\vbox{\settabs 11\columns
\toprule
\+\hfil face&equation&\hfil$L$&\hfil$\Nuz$&\hfil$E$&\hfil$B$&\hfil$C$&\hfil$D$&\hfil$p$&\cr
\midrule
\+\hfil up&\eqref{eqn:upward}~$\Nuols$&\hfil$\Ls$&\hfil$\Nuzs$&\hfil1&\hfil$2$&\hfil$1/\sqrt8$&\hfil$1$&\hfil$1/2$&\cr
\+\hfil vertical&\eqref{eqn:vertical}~$\Nuolq$&\hfil$L'$&\hfil$\Nuzq$&\hfil1&\hfil$1/2$&\hfil$1/2$&\hfil$1/4$&\hfil$1/2$&\cr
\+\hfil down&\eqref{eqn:downward}~$\Nuol_R$&\hfil$L_R$&\hfil$\Nuzq/2$&\hfil3&\hfil$4$&\hfil$1/2$&\hfil$2$&\hfil$1$&\cr
\bottomrule
}

\section{Self-Obstruction of Vertical and Downward-Facing Plates}

  Defined in formula~\eqref{eqn:variable-groups}, the Prandtl
  number~($\Pra$) is the momentum diffusivity per thermal diffusivity
  ratio of a fluid.  Heat transfer from plates to fluids with small
  $\Pra$ is primarily conduction.  Temperature changes in fluids with
  large~$\Pra$ cause changes in density which induce fluid flow that
  transports heat.  The~$\Nuolq$ and $\Nuol_R$ upper-bound
  formulas~\eqref{eqn:vertical<} and~\eqref{eqn:down<} are asymptotic for
  large~$\Pra$.

  Other than as a factor of~$\Ra$, $\Pra$ does not affect
  upward-facing heat transfer because the heated fluid flows directly
  upward, as does conducted heat.  When heated fluid must take longer
  paths around self-obstructing vertical and downward-facing plates,
  its heat transfer potential is reduced.

  $\Ra$ scales with $L^3$; $\Pra$ is a property of 3-dimensional
  fluids.  A function of $\Pra$ having values between~0 and~1 should
  scale~$\Ra$ in the vertical and downward-facing formulas.
  Both Schulenberg~\cite{SCHULENBERG1985467} and Churchill and
  Chu~\cite{churchill1975correlating} realized their formulas'
  dependence on~$\Pra$ in this way.
  Expressing the denominator of vertical formula~\eqref{eqn:Churchill-Chu}
  as the sixth root of an $\ell^p$-norm expression named $\Xi'(\Pra)$:
$$\left[1+{\left(0.492/\Pra\right)^{9/16}}\right]^{8/27}
 =\left[{\left\|1~,~{0.492\over\Pra}\right\|_{9/16}}\right]^{1/6}
 =\left[\Xi'(\Pra)\right]^{1/6}\eqdef{eqn:Xi'1}$$

  Scaling $\Ra$ by $1/\,\Xi'(\Pra)$ in the vertical upper-bound
  formula~\eqref{eqn:vertical<} makes it equivalent to
  formula~\eqref{eqn:Churchill-Chu}:
$$\Nuol^{1/2}=0.826+0.387\left[{\Ra\over\Xi'(\Pra)}\right]^{1/6}
 \quad\Xi'(\Pra)={\left\|1~,~{0.492\over\Pra}\right\|_{9/16}}\eqdef{eqn:Xi'}$$

  Similar treatment of the Schulenberg downward-facing strip and disk
  formulas \eqref{eqn:Schulenberg-strip} and \eqref{eqn:Schulenberg-disk}
  yields:
$$\eqalignno{\Nuol=0.544\left[{\,\Ra\over\Xi_R(\Pra)}\right]^{1/5}
 \qquad\Xi_R(\Pra)&=\left\|1~,~{0.477\over\Pra}\right\|_{3/5}&\eqdef{eqn:Xi_R}\cr
 \Nuol=0.619\left[{\,\Ra\over\Xi_r(\Pra)}\right]^{1/5}
 \qquad~\,\Xi_r(\Pra)&=\left\|1~,~{0.520\over\Pra}\right\|_{3/5}&\eqdef{eqn:Xi_r}\cr}$$

$$\lim_{\Pra\to+0}{1\over\Xi(\Pra)}=0\qquad\lim_{\Pra\to+\infty}{1\over\Xi(\Pra)}=1\eqdef{eqn:limits}$$


  The functions $\Xi'(\Pra)$, $\Xi_R(\Pra)$, and $\Xi_r(\Pra)$ are
  quite similar.  Coefficients 0.492, 0.477, and 0.520 are all within
  5\% of 1/2.  This suggests using 1/2 as the coefficient in a unified
  function $\Xi_\forall(\Pra)$.  The $\ell^p$-norm $p$ parameters
  $3/5=0.6$ and $9/16=0.5625$ differ by less than~7\%; and
  $\sqrt{1/3}\approx0.577$ lies between them:
$$\Xi_\forall(\Pra)={\left\|1~,~{0.5\over\Pra}\right\|_{\sqrt{1/3}}}\eqdef{eqn:Xi}$$

  Incorporating $\Xi_\forall(\Pra)$ into $\Nuolq$ and $\Nuol_R$
  upper-bound formulas~\eqref{eqn:vertical<} and~\eqref{eqn:down<} creates
  comprehensive vertical and downward convection
  formulas~\eqref{eqn:vertical} and~\eqref{eqn:downward}:
$$\eqalignno{\Nuolq=\left\|{{\Nuzq}\over2}~,~{\Nuzq^{4/3}\over8\,\root3\of2}\left[{\Ra\over\Xi_\forall(\Pra)}\right]^{1/3}\right\|_{1/2}
  &\approx\left\|0.682~,~0.150\left[{\Ra\over\Xi_\forall(\Pra)}\right]^{1/3}\right\|_{1/2}&\eqdef{eqn:vertical}\cr
  \Nuol_R=\left\|{\Nuzq\over4}~,~{\Nuzq^{6/5}\over2^{7/5}}\left[{\Ra\over\Xi_\forall(\Pra)}\right]^{1/5}\right\|_1
  ~~\,&\approx~\,0.341+0.550\left[{\Ra\over\Xi_\forall(\Pra)}\right]^{1/5}&\eqdef{eqn:downward}\cr}$$






\section{$\Ra$ Scaling Factors}

\vbox{\settabs 1\columns
\+\hfil\figscale{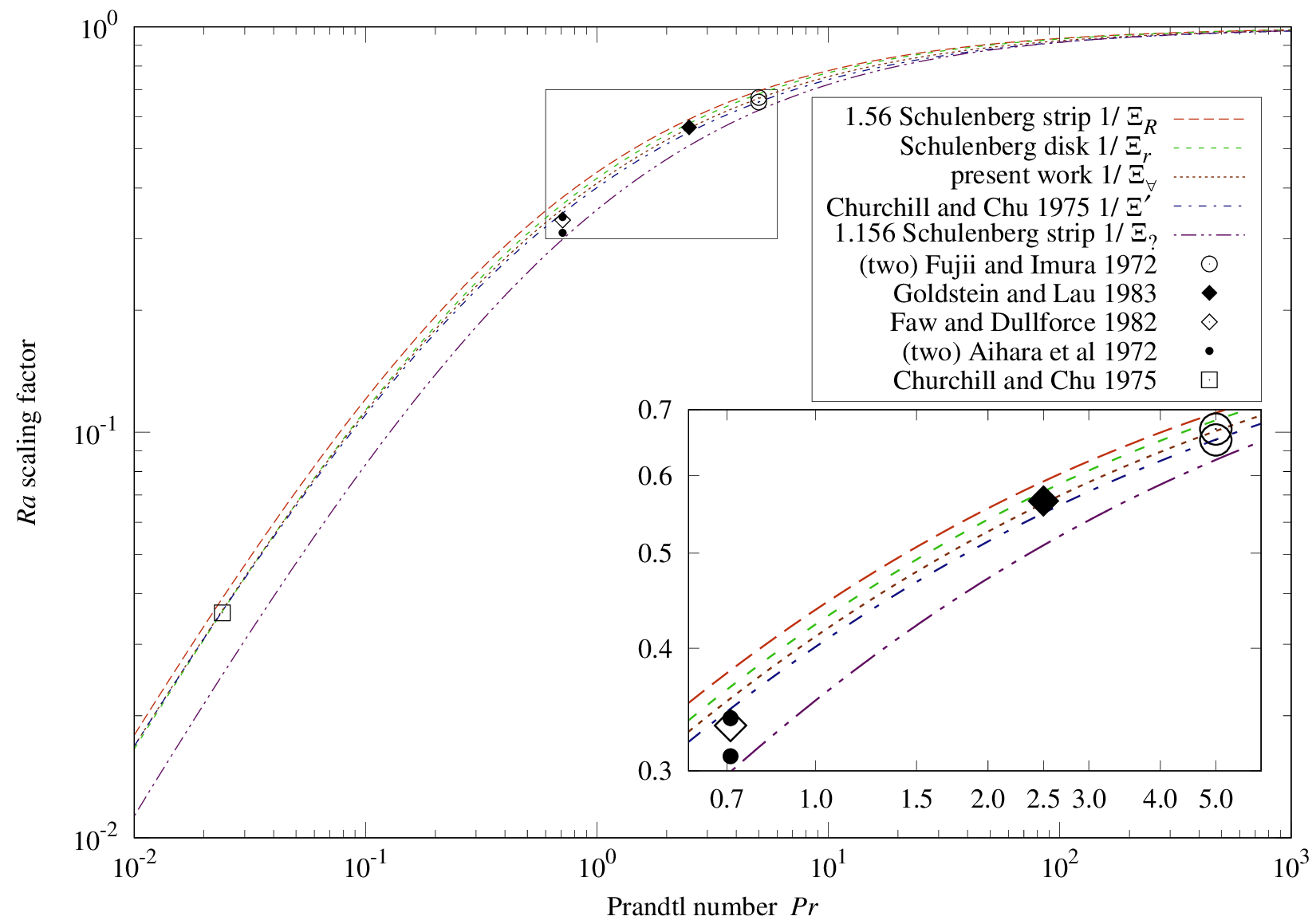}{468pt}&\cr
\+\hfil{\bf\figdef{fig:Ra-factor}\quad $\Ra$ scaling functions}&\cr
}

\smallskip
\vbox{\settabs 1\columns
\+\hfil{\bf\tabdef{tab:Ra-factor-tab}\quad{$\Ra$ scaling factors}}&\cr
\+\vbox{\settabs 10\columns
\toprule
\+face&\hfill source&&\hfill$\Pra$&\quad~fitted $1/\Xi$&\hfill$1/\,\Xi_R$~&\hfill$1/\,\Xi_r$~&\hfill$1/\,\Xi_\forall$~&\hfill$1/\,\Xi'$~&\hfill$1/\,\Xi_?$~&\cr
\midrule
\+vertical&Fujii and Imura~\cite{fujii1972natural}&&\hfill 5.00&\hfill 0.669&\hfill -3.7\%&\hfill -2.0\%&\hfill +0.5\%&\hfill +2.5\%&\hfill +7.5\%&\cr
\+down&Fujii and Imura~\cite{fujii1972natural}&&\hfill 5.00&\hfill 0.652&\hfill -6.1\%&\hfill -4.5\%&\hfill -2.1\%&\hfill -0.1\%&\hfill +4.8\%&\cr
\+down&Goldstein and Lau~\cite{goldstein_lau_1983}&&\hfill 2.50&\hfill 0.565&\hfill -4.5\%&\hfill -2.2\%&\hfill +0.6\%&\hfill +2.9\%&\hfill +10.9\%&\cr
\+down&Faw and Dullforce~\cite{FAW19821157}&&\hfill 0.71&\hfill 0.334&\hfill -12.1\%&\hfill -8.7\%&\hfill -6.2\%&\hfill -3.9\%&\hfill +11.5\%&\cr
\+down&Aihara et al~\cite{AIHARA19722535}&&\hfill 0.71&\hfill 0.339&\hfill -10.7\%&\hfill -7.2\%&\hfill -4.7\%&\hfill -2.3\%&\hfill +13.3\%&\cr
\+down&Aihara et al~\cite{AIHARA19722535}&&\hfill 0.71&\hfill 0.310&\hfill -18.4\%&\hfill -15.2\%&\hfill -12.8\%&\hfill -10.7\%&\hfill +3.6\%&\cr
\+vertical&Churchill and Chu~\cite{churchill1975correlating}&&\hfill 0.024&\hfill 0.036&\hfill -7.9\%&\hfill -0.8\%&\hfill -1.5\%&\hfill -0.9\%&\hfill +42.4\%&\cr
\bottomrule
}&\cr
}
\smallskip

  \figref{fig:Ra-factor} presents the $\Ra$ scaling functions
  $1/\,\Xi(\Pra)$ used in the vertical and downward-facing formulas.
  Also shown are~$1/\,\Xi(\Pra)$ values calculated from
  coefficients fitted to measurements in the cited sources.

  \tabref{tab:Ra-factor-tab} lists each fitted $1/\,\Xi$ value and its
  deviation from each $\Ra$ scaling function at the given~$\Pra$.
  Churchill and Chu~$1/{\Xi'}$ is closest to the fitted values.


  If a candidate formula is correct, negative deviations of fitted
  (aggregate) values can result from flow obstructions or measurement
  bias; positive deviations can result only from measurement bias.

  The positive $1/{\Xi'}$ deviations, $+2.5\%$ and $+2.9\%$, would
  indict two of the five sources in \tabref{tab:Ra-factor-tab}.

  The positive $1/\,\Xi_\forall$ deviations, $+0.5\%$~and $+0.6\%$,
  are tolerable.

  With values between those of~$1/\,\Xi_R$ and~$1/\,\Xi'$,
  $1/\,\Xi_\forall$~is the most plausible of these $\Ra$ scaling functions.


\medskip

  The ``1.156 Schulenberg strip $1/\,\Xi_?$'' curve
  in \figref{fig:Ra-factor} is the $\Ra$ scaling function from
  formula~\eqref{eqn:Schulenberg-strip-orig}.  \ref{Prior Work} argues
  that this formula from Schulenberg~\cite{SCHULENBERG1985467}
  contained a typographical error.  Being 40\% less than the other
  curves at $\Pra\ll1$ corresponds to
  formula~\eqref{eqn:Schulenberg-strip-orig} taking values 10\% less than
  (corrected) formula~\eqref{eqn:Schulenberg-strip}.


\section{Vertical Measurements}

\vbox{\settabs 1\columns
\+\hfil\figscale{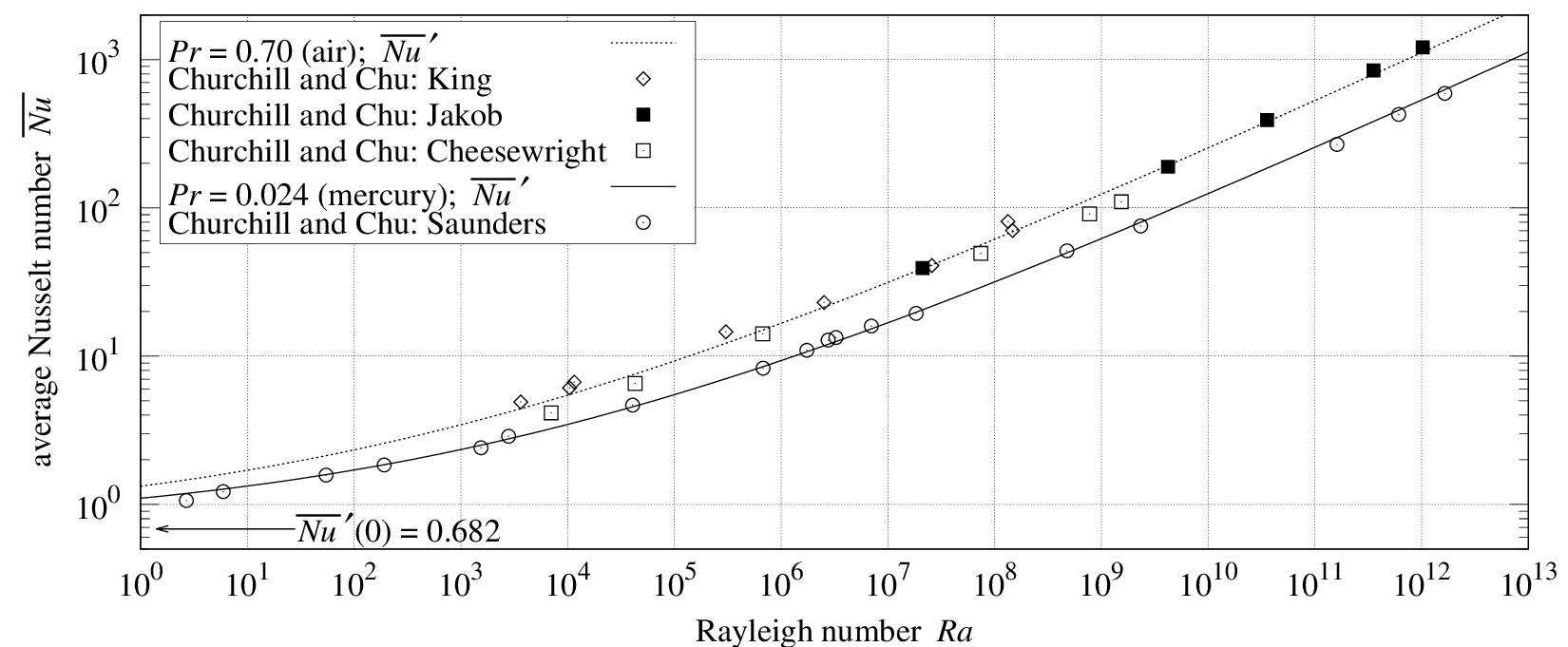}{468pt}&\cr
\+\hfil{\bf\figdef{fig:vertplate}\quad{heat transfer from vertical rectangular plate}}&\cr
}

\centerline{\bf\tabdef{tab:vert-conformance}\quad{heat transfer from vertical rectangular plate}}
\vbox{\offinterlineskip\settabs 9\columns
\toprule
\+\hfil source&&data-set&\hfil$\Pra$&\hfil face&\quad~formula&\hfil RMSRE&\quad bias\qquad scatter&\hfill count&\cr
\midrule
\+ Churchill and Chu~\cite{churchill1975correlating} -- King&&&\hfil 0.70&\hfil              vertical&\quad\eqref{eqn:vertical}~$\Nuolq$&\hfill 13.5\%~\quad&\hfill $+11.1\%$\qquad& 7.6\%\hfill 8&\cr
\+ Churchill and Chu~\cite{churchill1975correlating} -- Jakob&&&\hfil 0.70&\hfil             vertical&\quad\eqref{eqn:vertical}~$\Nuolq$&\hfill  4.7\%~\quad&\hfill $+3.1\%$\qquad& 3.5\%\hfill 5&\cr
\+ Churchill and Chu~\cite{churchill1975correlating} -- Cheesewright&&&\hfil 0.70&\hfil      vertical&\quad\eqref{eqn:vertical}~$\Nuolq$&\hfill 16.4\%~\quad&\hfill$-15.4\%$\qquad& 5.6\%\hfill 6&\cr
\+ Churchill and Chu~\cite{churchill1975correlating} -- Saunders&&&\hfil 0.024&\hfil         vertical&\quad\eqref{eqn:vertical}~$\Nuolq$&\hfill  5.3\%~\quad&\hfill $-1.2\%$\qquad& 5.1\%\hfill18&\cr
\bottomrule
}

  \figref{fig:vertplate} and \tabref{tab:vert-conformance} present four
  data-sets from Churchill and Chu~\cite{churchill1975correlating}.
  Differences between RMSRE computed from~$\Nuolq$
  formula~\eqref{eqn:vertical} and Churchill and Chu
  formula~\eqref{eqn:Churchill-Chu} are less than 1\%.

\section{Downward-Facing Measurements}

\vbox{\settabs 1\columns
\+\hfil\figscale{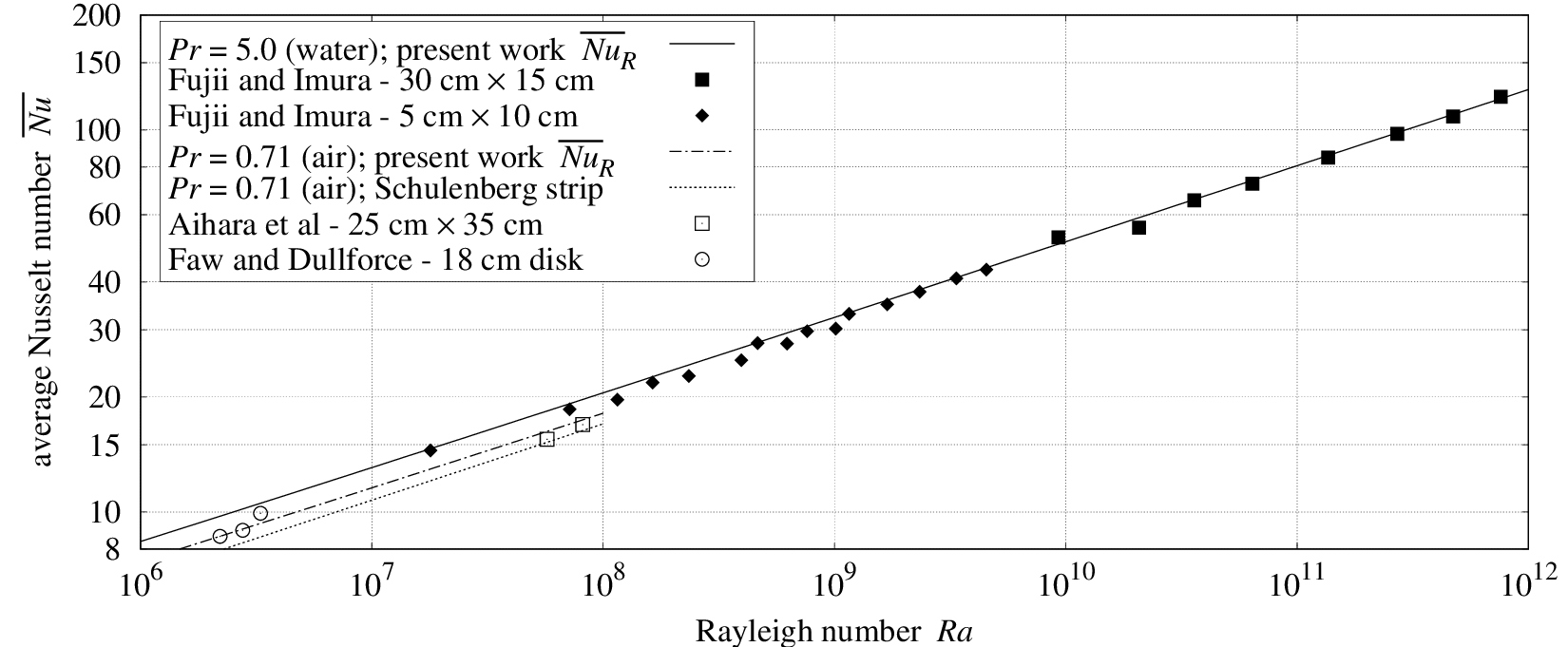}{468pt}&\cr
\+\hfil{\bf\figdef{fig:down}\quad{downward convection heat transfer from horizontal plate}}&\cr
}

\centerline{\bf\tabdef{tab:down-conformance}\quad{downward convection heat transfer from horizontal plate}}
\vbox{\offinterlineskip\settabs 9\columns
\toprule
\+\hfil source&&data-set&\hfil$\Pra$&\hfil face&\quad~formula&\hfil RMSRE&\quad bias\qquad scatter&\hfill count&\cr
\midrule
\+ Fujii and Imura~\cite{fujii1972natural} -- $30~\rm{cm}\times15~\rm{cm}$&&&\hfil 5.0&\hfil down&\quad\eqref{eqn:downward}~$\Nuol_R$&\hfill  2.7\%~\quad&\hfill $-0.8\%$\qquad& 2.6\%\hfill 8&\cr
\+ Fujii and Imura~\cite{fujii1972natural} -- $5~\rm{cm}\times10~\rm{cm}$&&&\hfil 5.0&\hfil  down&\quad\eqref{eqn:downward}~$\Nuol_R$&\hfill  4.2\%~\quad&\hfill $-3.4\%$\qquad& 2.5\%\hfill15&\cr
\+ Aihara et al~\cite{AIHARA19722535} -- $25~\rm{cm}\times35~\rm{cm}$&&&\hfil 0.71&\hfil  down&\quad\eqref{eqn:downward}~$\Nuol_R$&\hfill  3.8\%~\quad&\hfill $-3.7\%$\qquad& 0.9\%\hfill 2&\cr
\+ Faw and Dullforce~\cite{FAW19821157} -- 18.1~cm disk&&&\hfil 0.71&\hfil  down&\quad\eqref{eqn:downward}~$\Nuol_R$&\hfill  3.7\%~\quad&\hfill $+1.8\%$\qquad& 3.2\%\hfill 3&\cr
\bottomrule
}


\section{Characteristic-Length Metrics}

  For upward convection, $\Ls$ (area-to-perimeter ratio) is
  well-defined for any flat, convex plate.

  The formulas for vertical~\eqref{eqn:vertical} and downward
  convection~\eqref{eqn:downward} were developed for rectangular plates.
  More general characteristic-length metrics are needed to
  model heat transfer from other plate shapes.

  The next sections develop new characteristic-length metrics for
  vertical and downward-facing plates, and re-derive the formulas
  for~$\Nuols$~\eqref{eqn:upward}, $\Nuolq$~\eqref{eqn:vertical}, and
  $\Nuol_R$~\eqref{eqn:downward} from alternative plate shapes.

\section{Downward-Facing Circular Plate}

  Schulenberg equations~\eqref{eqn:Schulenberg-strip}
  and~\eqref{eqn:Schulenberg-disk} for downward-facing strips and disks
  have matching exponents, but their coefficients and
  characteristic-lengths differ: $L_R$~is $1/2$ of the rectangle's
  shorter side, versus disk radius $R$.  Can~$\Nuol_R$
  formula~\eqref{eqn:downward} predict heat transfer for both shapes using
  a single characteristic-length metric?

\subsection{Harmonic Mean}
  \figref{fig:below-flow}
  shows fluid closest to the heated surface flowing outward from the
  mid-line.
  If the plate edge is at varying distances from the mid-line (in the
  direction of flow), then some sort of length averaging is needed.
  Flow will be faster over the shorter distances because it
  experiences less drag; this suggests use of the ``harmonic~mean'',
  in which small values have more influence than large.

  For rectangles and disks, the mid-line is one of the plate's
  equal-area bisectors.  For rectangles it is parallel to the longer
  sides; but it is not the longest bisector, which is a diagonal.
  Being parallel to the longer sides implies that the mid-line is
  perpendicular to the shorter sides.  It will also be perpendicular
  to the shortest equal-area bisector; and this works for disks as
  well, where all diameters are bisectors.

  Consider a flat heated surface with its convex perimeter defined by
  functions $y_+(x)>0$ and $y_-(x)<0$ within the range $-R<x<R$ along
  the equal-area bisector which is perpendicular to the shortest
  equal-area bisector.
  Let $L_R$ be the combined harmonic mean of~$|y_+(x)|$
  and~$|y_-(x)|$:
$$L_R=1\left/\int_{-R}^R\left[{1\over |y_+(x)|}+{1\over |y_-(x)|}\right]\,{\diff{x}\over4\,R}\right.\eqdef{eqn:L_R}$$

  For rectangular plates, $L_R$ formula~\eqref{eqn:L_R} is $1/2$ of the
  shorter side's length, the same characteristic-length used by
  Schulenberg~\cite{SCHULENBERG1985467}.  For disks of radius~$R$:
$$y_+(x)=-y_-(x)=\sqrt{R^2-x^2}\qquad
  L_R=4\,R\left/\int_{-R}^R{2\over\sqrt{R^2-x^2}}\,{\diff{x}}\right.={2\over\pi}\,R\eqdef{eqn:L_R=}$$

  Schulenberg's disk formula~\eqref{eqn:Schulenberg-disk} used radius~$R$
  as the disk's characteristic-length.
  Converting the~$\Nuol_r$
  coefficient from characteristic-length~$R$ to~$L_R$ is
  $0.619\,[{2/\pi}]^{1/5}\approx0.566$, which is within 3\% of the
  0.550 coefficient of $\Nuol_R$ formula \eqref{eqn:downward}.

\subsection{Recalculate Conduction}
  The derivation of downward
  formulas~(\eqrefn{eqn:down<},~\eqrefn{eqn:downward}) was for a square plate.
  To derive the downward formula for a disk, $\Nuz$ will be
  recalculated; it will be scaled larger because of the increased flow
  over the shorter distances.  Recalling from \ref{Upward-Facing
  Circular Plate},
  the conduction shape factor for one side of a disk is $S=2\,D=4\,R$.
$$\Nuz={S\,L_R\over A}={4\,R\,L_R\over\pi\,R^2}={8\over\pi^2}\approx0.811\eqdef{eqn:disk-down}$$

  Upward-facing disk $\Nuzs$ was scaled by $\sqrt{1/8}$ in
  formula~\eqref{eqn:I_p=*} because its flow was midway between parallel
  and radially inward; downward-facing $\Nuzq/2$ was unscaled in
  (rectangle) formula~\eqref{eqn:I_p=R}.  The flow from a downward-facing
  disk spreads from a diameter; an intermediate scale is needed.  The
  geometric mean of 1~(unscaled) and $\sqrt{1/8}$ is $\root4\of{1/8}$.
  Scaling~$\Nuz$ by the reciprocal, $\root4\of{8}\approx1.682$, yields
  the rectangular~$\Nuzq\approx1.363$.
  With the $L_R$ harmonic mean metric \eqref{eqn:L_R} and
  ${\root4\of{8}\,\Nuz}=\Nuzq$, the rest of the derivation is
  unchanged.  Thus, downward formula~\eqref{eqn:downward} works for both
  rectangular and circular plates.

\subsection{Measurements}
  \figref{fig:down} and \tabref{tab:down-conformance} include three disk
  measurements from Faw and Dullforce~\cite{FAW19821157}.

\subsection{Rectangle Shape Factor}
  ${\root4\of{8}\,\Nuz}=\Nuzq$ derives an exact expression for the
  rectangular plate's dimensionless shape factor~$q_{SS}^*$:
$${q_{SS}^*}^2\,{\pi\over2}={8\,\root4\of8\over\pi^2}
  \qquad{q_{SS}^*}={4\,\root8\of8\over\pi^{3/2}}
  \approx0.93158+\eqdef{eqn:q_s*}$$


\section{Vertical Circular Plate}

  Consider a flat vertical plate as an array of narrow vertical
  plates.
  With~$\Ra=0$, integrate ${h}$ across the vertical slices of
  heights~$L(x)$; then solve $\overline{h}={\Nuzq\,k/L'}$ for~$L'$:
$${\Nuzq\,k\over L'}=\overline{h}={1\over2\,R}\int_{-R}^R{\Nuzq\,k\over L(x)}\,\diff{x}
 \qquad L'=2\,R\left/{\int_{-R}^R{1\over L(x)}\,\diff{x}}\right.\eqdef{eqn:L'}$$

  Therefore, the vertical characteristic-length $L'$ is the harmonic
  mean of vertical spans $L(x)$.

  For a rectangle of height $L$, $L'=L$.  For the disk of radius~$R$,
  $L'=4\,R/\pi$. Recalculating~$\Nuz$:
$$\Nuz={S\,L'\over A}={4\,R\,L'\over\pi\,R^2}={16\over\pi^2}\approx1.621\eqdef{eqn:disk-vert}$$

  Instead of the bifurcated flow from the downward-facing plate, flow
  is along full vertical spans of the disk; 1/2 of the
  diameter-to-circle fringing which scaled downward-facing disk~$\Nuz$
  by~$\root4\of{8}$ should apply to vertical disks.  With this
  scaling, $\Nuz\,\root4\of{8}/2=\Nuz/\root4\of{2}=\Nuzq\approx1.363$,
  and the rest of the derivation is unchanged.  Thus, vertical
  formula~\eqref{eqn:vertical} works for disks and axis-aligned
  rectangular plates using $L'$ harmonic mean formula~\eqref{eqn:L'}.

\subsection{Measurements}
  Kobus and Wedekind~\cite{KOBUS19953329} measured natural convection
  heat transfer from vertical thermistor disks in air.  They also
  included a series of similar measurements from Hassani and
  Hollands~(1987).  $\Pra$~was not specified; 0.71 is assumed.  Each
  disk's diameter~$d$ and thickness~$t$ are specified
  in \figref{fig:vertdisk}.

  Air heated by the two disk sides flows over the upper half of the
  rim, inhibiting upper rim heat transfer.  The lower half of the rim
  transfers heat at the same per area rate as the sides.  The disk
  effective surface area, including half of the rim, is
  $\pi\,d^2/2+\pi\,d\,t/2$.  Being normalized for area, $\Nuolq$
  should be scaled by:
$${\pi\,d^2/2\over\pi\,d^2/2+\pi\,d\,t/2}={d^2\over d^2+d^2\,[t/d]}={1\over1+t/d}\eqdef{eqn:rim}$$

  To convert characteristic-lengths to the harmonic mean, Kobus and
  Wedekind~$\Ra$ gets scaled by $[2/\pi]^3$; $\Nuol$ gets scaled
  by~$[2/\pi]$ and formula~\eqref{eqn:rim}.  Hassani and Hollands used
  different characteristic-lengths for~$\Ra$ and~$\Nuol$; their~$\Ra$
  gets scaled by~$1/[2\,\sqrt{\pi}]^3$, while~$\Nuol$ gets scaled
  by~$\sqrt{\pi}/4$ and formula~\eqref{eqn:rim}.

\smallskip
\vbox{\settabs 1\columns
\+\hfil\figscale{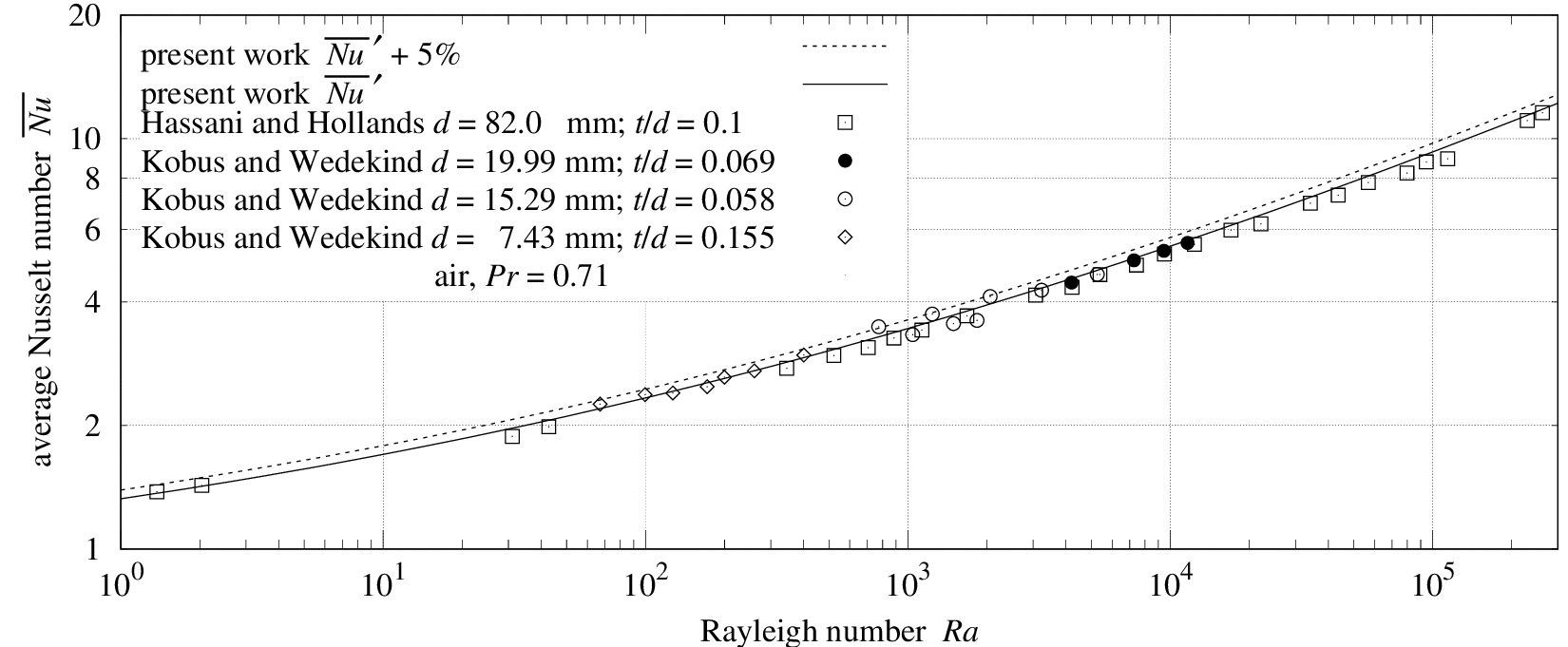}{468pt}&\cr
\+\hfil{\bf\figdef{fig:vertdisk}\quad{heat transfer from vertical disk}}&\cr
}

\smallskip
\centerline{\bf\tabdef{tab:vertdisk-conformance}\quad{heat transfer from vertical disk}}
\vbox{\settabs 9\columns
\toprule
\+\hfil source&&data-set&\hfil$\Pra$&\hfil face&\quad~formula&\hfil RMSRE&\quad bias\qquad scatter&\hfill count&\cr
\midrule
\+ Hassani and Hollands~\cite{KOBUS19953329} -- 82 mm&&&\hfil 0.71&\hfil                vertical disk&\quad\eqref{eqn:vertical}~$\Nuolq$&\hfill  3.8\%~\quad&\hfill $-3.4\%$\qquad& 1.6\%\hfill26&\cr
\+ Kobus and Wedekind~\cite{KOBUS19953329} -- three sizes&&&\hfil 0.71&\hfil            vertical disk&\quad\eqref{eqn:vertical}~$\Nuolq$&\hfill  3.2\%~\quad&\hfill $-0.4\%$\qquad& 3.1\%\hfill19&\cr
\bottomrule
}

  \figref{fig:vertdisk} and \tabref{tab:vertdisk-conformance} present the
  vertical disk heat transfer measurements.

  In the clever design using thermistors, only horizontal clearance
  and the wires attached to the disk centers remained as obstructions,
  achieving a close match with the present theory
  in \tabref{tab:vertdisk-conformance}.

\section{Upward-Facing Square Plate}

  Converting square plate~$\Nuz$ formula~\eqref{eqn:Nu_0} from $L'$
  to~$\Ls$ is division by 4.
  Expanding~${q_{SS}^*}$ from equation~\eqref{eqn:q_s*}:
$${\Nuz\over4}={{q_{SS}^*}\over4}\,{\sqrt{\pi\over2}}={\root8\of8\over\sqrt{2}\,\pi}={8^{5/8}\over4\,\pi}\approx0.292\eqdef{eqn:square*}$$

  The square plate's flow will be moderately radial, scaling midway
  between the reciprocal of~$\sqrt{1/8}$ from the upward-facing disk,
  and~$\root4\of{8}$ from the downward-facing disk's bifurcated flow.
  The geometric mean of~$\sqrt{8}$ and~$\root4\of{8}$
  is~$8^{3/8}\approx2.181$.  Scaling $\Nuz/4$ by $8^{3/8}$ yields
  $2/\pi=\Nuzs\approx0.637$ from equation~\eqref{eqn:Nu0*}.
  The rest of the derivation is unchanged.  Thus, upward convection
  $\Nuols$ formula~\eqref{eqn:upward} works for disks and square plates.

\section{Vertical Rectangular Plate With Side-Walls}

 The Fujii and Imura~\cite{fujii1972natural} apparatus had (unheated)
 perpendicular side-walls of length~$L'$ forming a channel with the
 plate.
 In \tabref{tab:vertical-with-sides}, the vertical 30~cm and 5~cm plates
 average 44\% and 18\% less heat transfer than expected by~$\Nuolq$
 vertical formula~\eqref{eqn:vertical}.  Clearly, formula~\eqref{eqn:vertical}
 is incorrect for vertical plates with side-walls.

 Formula~\eqref{eqn:V-I_k1} incorporates an additional factor of~$\Rey$
 in~$I_k$ to model the side-wall drag.  The side-walls obstruct
 horizontal flow, so the combined length of flow along the plate and
 side-wall is~$2\,L$; heat transfer contact along the plate is~$L$.
 The $\ell^{1/2}$-norm changes to the $\ell^1$-norm
$$\eqalignno{I_k&={2\,\Rey}\,{\Rey}\,{\rho\,u^3\over2}\,{\Pi_4\over2}\,\Pi_5
     ={L^2\over\nu^2}\,{\rho\,u^5\over2\,\Pi_4\,\Pi_5}\qquad
 u=\left[{\nu^2\over L^2}\,{2\,I_k\over\rho\,\Pi_4\,\Pi_5}\right]^{1/5}&\eqdef{eqn:V-I_k1}\cr
 I_p&={k\,\Delta{T}\over L}\,\Nuzq\,\left\|{1\over2}~,~{1\over2}\,{\Rey}\right\|_1
     ={k\,\Delta{T}\over L}\,\Nuzq\,\left[{1\over2}+{L\over2\,\nu}\,\left[{\nu^2\over L^2}\,{2\,I_k\over\rho\,\Pi_4\,\Pi_5}\right]^{1/5}\right]&\eqdef{eqn:V-I_p1}\cr}$$





 Using the general formula~\eqref{eqn:general} with $E=2$, $B=2$, and
 $D=1$:
$$\Nuolq_w={\Nuzq\over2}+{\Nuzq^{5/4}\over2^{5/4}}\,\left[{\Ra\over\Xi_\forall(\Pra)}\right]^{1/4}\eqdef{eqn:Nu'_w}$$

\vbox{\settabs 1\columns
\+\hfil\figscale{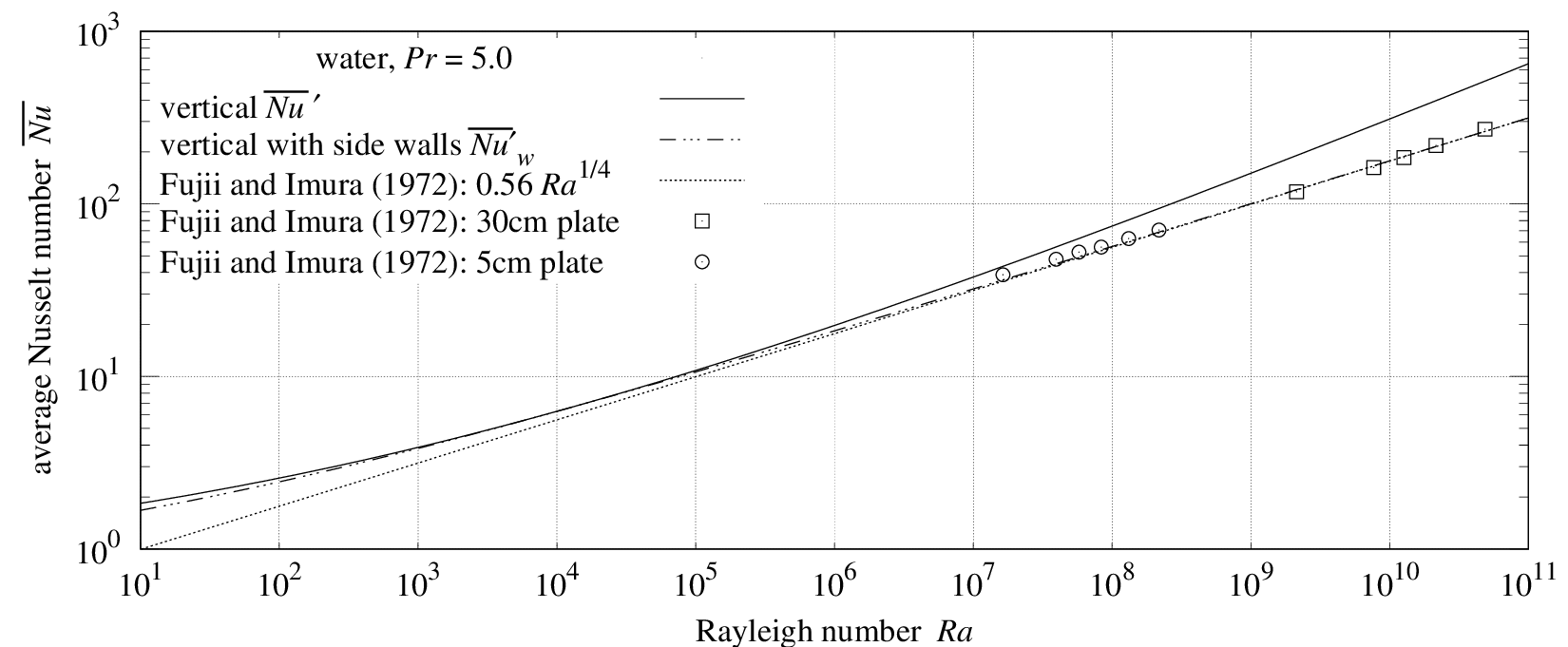}{468pt}&\cr
\+\hfil{\bf\figdef{fig:VSW}\quad vertical Fujii and Imura plates}&\cr
}
\medskip

\centerline{\bf\tabdef{tab:vertical-with-sides}\quad{vertical Fujii and Imura plates}}
\vbox{\settabs 9\columns
\toprule
\+\hfil source&&data-set&\hfil$\Pra$&\hfil face&\quad~formula&\hfil RMSRE&\quad bias\qquad scatter&\hfill count&\cr
\midrule
\+ Fujii and Imura~\cite{fujii1972natural} -- $30~\rm{cm}\times15~\rm{cm}$&&&\hfil 5.0&\hfil vertical&\quad\eqref{eqn:vertical}~$\Nuolq$&\hfill 43.8\%~\quad&\hfill $-43.7\%$\qquad& 3.0\%\hfill 5&\cr
\+ Fujii and Imura~\cite{fujii1972natural} -- $30~\rm{cm}\times15~\rm{cm}$&&&\hfil 5.0&\hfil vertical&\quad\eqref{eqn:Nu'_w}~$\Nuolq_w$&\hfill   2.2\%~\quad&\hfill  $-0.5\%$\qquad& 2.2\%\hfill 5&\cr
\+ Fujii and Imura~\cite{fujii1972natural} -- $5~\rm{cm}\times10~\rm{cm}$&&&\hfil 5.0&\hfil  vertical&\quad\eqref{eqn:vertical}~$\Nuolq$&\hfill 18.9\%~\quad&\hfill $-18.3\%$\qquad& 4.7\%\hfill 6&\cr
\+ Fujii and Imura~\cite{fujii1972natural} -- $5~\rm{cm}\times10~\rm{cm}$&&&\hfil 5.0&\hfil  vertical&\quad\eqref{eqn:Nu'_w}~$\Nuolq_w$&\hfill   5.2\%~\quad&\hfill  $+5.0\%$\qquad& 1.6\%\hfill 6&\cr
\bottomrule
}

 \tabref{tab:vertical-with-sides} details the performance of
 formula~\eqref{eqn:Nu'_w} for vertical plates with side-walls.  RMSRE of
 2.2\% and 5.2\% are substantial decreases from 43.8\% and
 18.9\%.  \figref{fig:VSW} shows $\Nuolq_w\le\Nuolq$, as required by
 efficiency constraints.

\section{Upward-Facing Rectangular Plate With Side-Walls}

  The $\Lsw$ area-to-perimeter ratio for side-walled upward-facing
  plates excludes side-wall length~$L'$ from the perimeter length
  ($2\,w+2\,L'$) because there is no flow through side-walls.  Thus,
  $\Lsw={L'\,w/[2\,w]}={L'/2}$.

 In \tabref{tab:up-conformance} (and \tabref{tab:upward-with-sides}), the
 Fujii and Imura data-sets averaged 0.4\% and 11.4\% less than
 expected from~$\Nuols$ formula~\eqref{eqn:upward}.  The 30~cm plate
 having side-walls twice as long as its 15~cm channel width made its
 flow similar to vertical plate flow in each half of the plate towards
 the center-line.  $L=\Lsw=L'/2$.

 Fluid rises after heating; so the $I_k$ and heat transfer factors are
 both $\Rey/2$.  As with the side-walled vertical plate, conduction
 and flow-induced heat transfer combine using the $\ell^1$-norm
 (addition):
$$\eqalignno{I_k&={\Rey\over2}\,{\rho\,u^3\over2}\,{\Pi_4\over2}\,\Pi_5
     ={L\over\nu}\,{\rho\,u^4\over8\,\Pi_4\,\Pi_5}\qquad
 u=\left[{\nu\over L}\,{8\,I_k\over\rho\,\Pi_4\,\Pi_5}\right]^{1/4}&\eqdef{eqn:U-I_k}\cr
 I_p&={k\,\Delta{T}\over L}\,{\Nuzq\over2}\,\left\|{1\over2}~,~{1\over2}\,{\Rey\over2}\right\|_1
     ={k\,\Delta{T}\over L}\,\Nuzq\,\left\|{1\over4}~,~{L\over8\,\nu}\,\left[{\nu\over L}\,{8\,I_k\over\rho\,\Pi_4\,\Pi_5}\right]^{1/4}\right\|_1
     &\eqdef{eqn:U-I_p}\cr}$$




 Flow is partially obstructed by the side walls; hence, $\Ra$~will be
 scaled by $1/\Xi(\Pra)$.  Using the general formula~\eqref{eqn:general}
 with $\Nuz=\Nuzq/2$, $B=1/2$, $C=1/2$, $D=1/2$, and $p=1$:
$$\Nuol_w={\Nuzq\over4}+{\Nuzq^{4/3}\over8\,\root3\of2}\,\left[\Ra'\over\Xi(\Pra)\right]^{1/3}\eqdef{eqn:U-Nu_w}$$

 Note that, unlike~$\Nuols$ formula~\eqref{eqn:upward},
 $\Nuol_w$~formula~\eqref{eqn:U-Nu_w} depends on $\Xi(\Pra)$.

\smallskip
\vbox{\settabs 1\columns
\+\hfil\figscale{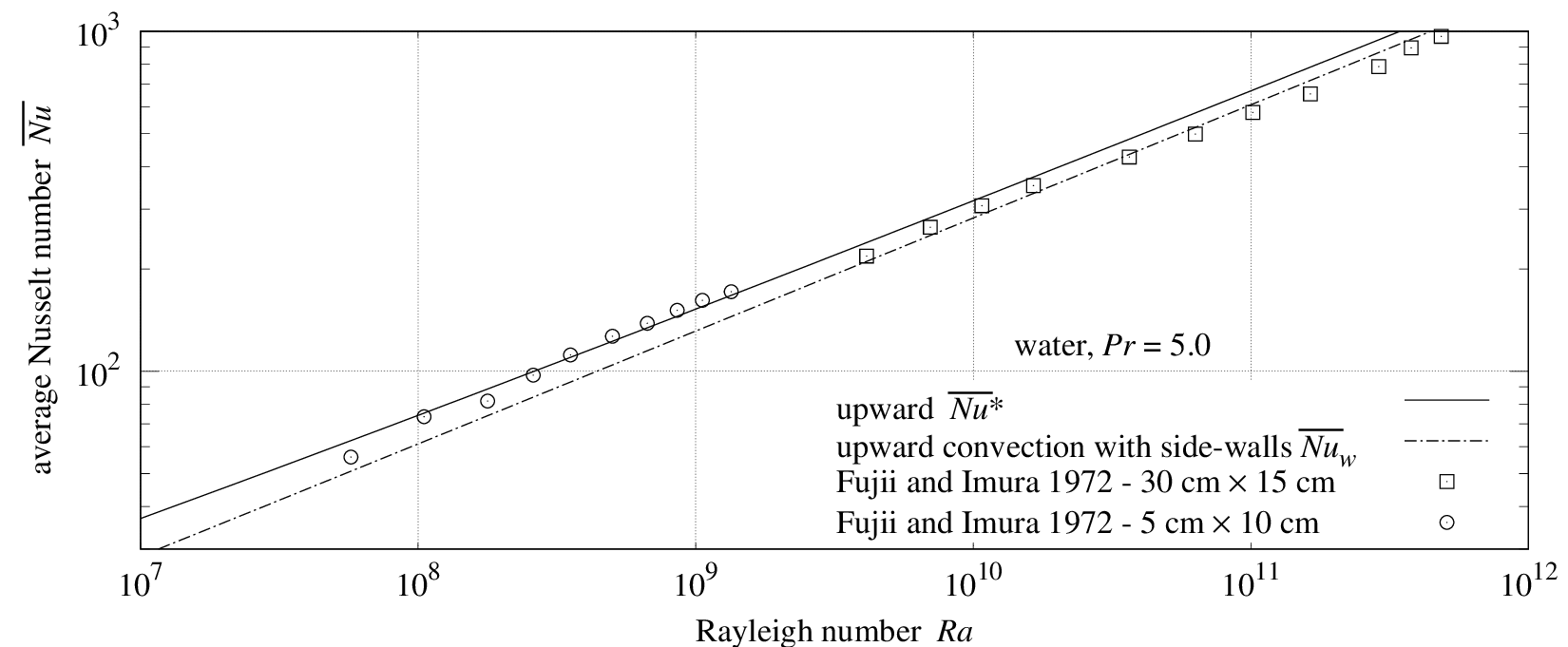}{468pt}&\cr
\+\hfil{\bf\figdef{fig:USW}\quad{upward-facing Fujii and Imura plates}}&\cr
}
\smallskip

\centerline{\bf\tabdef{tab:upward-with-sides}\quad{upward-facing Fujii and Imura plates}}
\vbox{\settabs 9\columns
\toprule
\+\hfil source&&data-set&\hfil$\Pra$&\hfil face&\quad~formula&\hfil RMSRE&\quad bias\qquad scatter&\hfill count&\cr
\midrule
\+ Fujii and Imura~\cite{fujii1972natural} -- $30~\rm{cm}\times15~\rm{cm}$&&&\hfil 5.0&\hfil   up&\quad\eqref{eqn:upward}~$\Nuols$&\hfill 12.0\%~\quad&\hfill$-11.4\%$\qquad& 4.0\%\hfill11&\cr
\+ Fujii and Imura~\cite{fujii1972natural} -- $30~\rm{cm}\times15~\rm{cm}$&&&\hfil 5.0&\hfil   up&\quad\eqref{eqn:U-Nu_w}~$\Nuol_w$&\hfill 6.0\%~\quad&\hfill$-1.8\%$\qquad& 5.7\%\hfill11&\cr
\+ Fujii and Imura~\cite{fujii1972natural} -- $5~\rm{cm}\times10~\rm{cm}$&&& \hfil 5.0&\hfil   up&\quad\eqref{eqn:upward}~$\Nuols$&\hfill  5.0\%~\quad&\hfill$-0.4\%$\qquad& 5.0\%\hfill10&\cr
\+ Fujii and Imura~\cite{fujii1972natural} -- $5~\rm{cm}\times10~\rm{cm}$&&& \hfil 5.0&\hfil   up&\quad\eqref{eqn:U-Nu_w}~$\Nuol_w$&\hfill  18.2\%~\quad&\hfill$+17.7\%$\qquad& 4.1\%\hfill10&\cr
\bottomrule
}
\smallskip

 \figref{fig:USW} and \tabref{tab:upward-with-sides} show that~$\Nuol_w$
 formula~\eqref{eqn:U-Nu_w} is effective for the 30~cm plate, but not for
 the 5~cm plate.
 The 5~cm plate's 10~cm channel width is twice the 5~cm side-wall
 length.  This makes its flow more radial than parallel.  Thus,
 $\Nuols$ formula~\eqref{eqn:upward} is more appropriate for the 5~cm
 plate.

 \tabref{tab:side-wall parameters} lists the general derivation
 formula\eqref{eqn:general} parameters for the side-wall flow topologies.

\smallskip
\centerline{\bf\tabdef{tab:side-wall parameters}\quad{side-walled plate parameters}}
\moveright 0.05\hsize\vbox{\settabs 10\columns
\toprule
\+\hfil face&equation&\hfil$L$&\hfil$\Nuz$&\hfil$E$&\hfil$B$&\hfil$C$&\hfil$D$&\hfil$p$&\cr
\midrule
\+\hfil up&\eqref{eqn:U-Nu_w}~$\Nuol_w$&\hfil$\Lsw=L'/2$&\hfil$\Nuzq/2$&\hfil1&\hfil$1/2$&\hfil$1/2$&\hfil$1/2$&\hfil$1$&\cr
\+\hfil vertical&\eqref{eqn:Nu'_w}~$\Nuolq_w$&\hfil$L'$&\hfil$\Nuzq$&\hfil2&\hfil$2$&\hfil$1/2$&\hfil$1$&\hfil$1$&\cr
\bottomrule
}


\section{Inclined Plate}

  Let $\theta$ be the angle of a plate from vertical.  Face up
  $\theta=-90^\circ$; vertical $\theta=0^\circ$; face down
  $\theta=+90^\circ$.  This investigation has derived and tested
  formulas for rectangular and circular plates at $-90^\circ$,
  $0^\circ$, and $+90^\circ$.


  Following Fujii and Imura~\cite{fujii1972natural}, the~$\Ra'$
  argument to vertical formula~\eqref{eqn:vertical} gets scaled
  by~${\left|\cos\theta\right|}$ to model the reduced vertical
  convection from an inclined plate.\numberedfootnote{Fujii and
  Imura~\cite{fujii1972natural} credits B. R. Rich with the idea of
  scaling vertical~$\Ra$ by $|\cos\theta|$.  It can be thought of as a
  reduction in the effective gravitational acceleration~$g$, which
  scales~$\Ra$ linearly.}
  Following 
  Raithby and Hollands~\cite{rohsenow1998handbook}, the~$\Ra^*$
  and~$\Ra_R$ arguments to the upward and downward
  formulas \eqref{eqn:upward} and \eqref{eqn:downward} get scaled by
  ${\left|\sin\theta\right|}$.

  Raithby and Hollands take the maximum convective surface
  conductance~$\overline{h}=k\,\Nuol/L$,
  not~$\Nuol$, to determine the overall convection.  This is because
  $\Nuolq$, $\Nuols$, and $\Nuol_R$ have different
  characteristic-lengths, while $\overline{h}$ is independent of $L$.
  Taking the maximum
  asserts that the associated flow topologies are mutually exclusive.
  This extreme competition is the $\ell^\infty$-norm, which is
  equivalent to the $\max(\,)$ function when all its arguments are
  non-negative:
$$\overline{h}=k\cases{
 \max\left({\Nuolq({\left|\cos\theta\right|}\,\Ra')/L'}, {\Nuol_R\left({\left|\sin\theta\right|}\,\Ra_R\right)/L'}\right)&$\Delta{T}\,\sin\theta\ge0$\cr
 \max\left({\Nuolq({\left|\cos\theta\right|}\,\Ra')/L'}, {\Nuols({\left|\sin\theta\right|}\,\Ra^*)/\Ls}\right)&$\Delta{T}\,\sin\theta\le0$\cr}
 \eqdef{eqn:inc}$$

  Rayleigh numbers can be expressed in terms of (vertical plate)~$\Ra'$:
$$\eqalignno{\overline{h}&=k\,\max\left({\Nuolq({\left|\cos\theta\right|}\,\Ra')\over L'}, {1\over L_R}\,{\Nuol_R\left({\left|\sin\theta\right|}\,\Ra'\left[{L_R\over L'}\right]^3\right)}\right)
 \qquad\Delta{T}\,\sin\theta\ge0&\eqdef{eqn:inc-down}\cr
 \overline{h}&=k\,\max\left({\Nuolq({\left|\cos\theta\right|}\,\Ra')\over L'}, {1\over \Ls}\,{\Nuols\left({\left|\sin\theta\right|}\,\Ra'\,\left[{\Ls\over L'}\right]^3\right)}\right)
 \qquad~\Delta{T}\,\sin\theta\le0&\eqdef{eqn:inc-up}\cr}$$

  The downward-facing topology flows outward from opposite plate
  edges; there is no plate area available for the vertical flow
  topology.  Thus, $\Nuolq$ and $\Nuol_R$ are mutually exclusive in
  formula~\eqref{eqn:inc-down}.  The upward-facing topology draws fluid
  from the whole perimeter.  Thus, $\Nuolq$ and $\Nuols$ are mutually
  exclusive in formula~\eqref{eqn:inc-up}.  Note that
  formula~\eqref{eqn:inc-up} mutual exclusion may not hold when part of
  the perimeter flow is obstructed.

  Ideally, when $\Ra=0$, $\overline{h}$~should be independent
  of~$\theta$.  For a 1~m square plate in~$k=1$ fluid,
  $\overline{h}(0^\circ)=\overline{h}(+90^\circ)=\Nuz'/2\approx0.682$,
  but $\overline{h}(-90^\circ)\approx1.646$.  When $\theta=-90^\circ$
  forces $\Ra'\,\cos\theta$ to~0, only the conduction term remains.
  As noted in \ref{Upward-Facing Measurements}, $\Nuols$
  formula~\eqref{eqn:upward} does not extend to static conduction.
  
  $\Nuols$ and $\Nuolq$ track measurements well near~$\Ra\approx1$
  in \figrefs{fig:nuhup}, \figrefn{fig:vertplate}, and \figrefn{fig:vertdisk}.
  Ignoring $\Nuols$ when $\Ra'\sin\theta>-[\Ls/L']^3$, and $\Nuol_R$
  when $\Ra'\sin\theta<[L_R/L']^3$, avoids the conduction term
  competition at~$\theta\approx0$:
$$\overline{h}=k\cases{
  \max\left({\Nuolq({\left|\cos\theta\right|}\,\Ra')/L'}, {\Nuols\left({\left|\sin\theta\right|}\,\Ra'\,[{\Ls/L'}]^3\right)/\Ls}\right)&$\Ra'\sin\theta<-[\Ls/L']^3$\cr
  \max\left({\Nuolq({\left|\cos\theta\right|}\,\Ra')/L'}, {\Nuol_R\left({\left|\sin\theta\right|}\,\Ra'\,[{L_R/L'}]^3\right)/L_R}\right)&$\Ra'\sin\theta>[L_R/L']^3$\cr
  {\Nuolq\left({\left|\cos\theta\right|}\,\Ra'\right)/L'}&otherwise.\cr
}\eqdef{eqn:h}$$
  Note that ${\Ls/L'}\le{1/2}$ and ${L_R/L'}\le{1/2}$ are true for any
  flat, convex plate face.

  For~$\Ra>1$, the proposed $\overline{h}$ formula~\eqref{eqn:h} will
  match non-side-walled horizontal and vertical plate measurements to
  their appropriate $k\,\Nuols\!/\Ls$, $k\,\Nuolq\!/L'$, and
  $k\,\Nuol_R/L_R$ values.


\section{Inclined Plate With Side-Walls}

  The Fujii and Imura~\cite{fujii1972natural} apparatus had
  length~$L'$ side-walls.  In \tabref{tab:down-conformance}, $\Nuol_R$
  formula~\eqref{eqn:downward} has less than~5\% RMSRE;
  it is used as the side-walled downward-facing formula
  with $L_R=L'/2$, regardless of which side is shorter.

  To adapt~$\overline{h}$ formulas \eqref{eqn:inc-down} and \eqref{eqn:inc-up}
  to apparatus with side-walls, $\Nuolq_w$~formula~\eqref{eqn:Nu'_w}
  replaces $\Nuolq$.  This leads to proposed downward~$\overline{h}$
  formula~\eqref{eqn:tilt-down} for side-walled plates:
$$\overline{h}=k\,\max\left({\Nuolq_w({\left|\cos\theta\right|}\,\Ra')\over L'}, {\Nuol_R\left({\left|\sin\theta\right|}\,\Ra'/2^3\right)\over L'/2}\right)\qquad\Delta{T}\,\sin\theta\ge0\eqdef{eqn:tilt-down}$$

  Fujii and Imura photographs show the plume originating in the middle
  of the 5~cm plate when $\theta=-90^\circ$.  The origin shifts 9\%
  toward the elevated end of the plate when $\theta=-85^\circ$.  The
  plume for the 30~cm plate at $\theta=-60^\circ$ originates in the
  upper 1/4 of the plate.  Plume movement with $\theta$ indicates that
  regions of upward-facing and vertical convection shared the
  side-walled plate in the Fujii and Imura apparatus.

\vfill\eject

  The side-walled upward and vertical topologies compete for
  horizontal flow in the channel.  If competition were between
  perpendicular flows, they would combine as the root-sum-squared,
  which is the $\ell^2$-norm.  To compete for horizontal channel flow,
  two changes in direction are required; they combine as the
  $\ell^4$-norm.

  The 30~cm plate has side-walls twice as long as its channel
  width~$w$.  Flow through this channel will be primarily parallel, as
  modeled by~$\Nuol_w$ formula~\eqref{eqn:U-Nu_w}.  The proposed upward
  $\overline{h}$ formula for the 30~cm plate is:
$$\overline{h}=k\,\left\|{\Nuolq_w({\left|\cos\theta\right|}\,\Ra')\over L'}~,~{{\Nuol_w\left({\left|\sin\theta\right|}\,\Ra'/2^3\right)}\over L'/2}\right\|_4\qquad\Delta{T}\,\sin\theta\le0\qquad w<L'\eqdef{eqn:tilt-up-30cm}$$

  The 5~cm plate's 10~cm channel is twice as wide as its length;
  horizontal flow will be more radial than parallel.  At
  $\theta=-90^\circ$ it is modeled by~$\Nuols$ formula~\eqref{eqn:upward},
  but with characteristic-length $\Lsw=L'/2$.

  Fujii and Imura streamlines photographs show the vertical $\Nuolq_w$
  and upward-facing $\Nuol_w$ flow modes having uniform rates of
  horizontal flow between the lower heated plate edge and pause
  elevation $z_t$.

  Uniform horizontal flow is not the case for the 5~cm plate at
  $\theta=-45^\circ$.  The horizontal flow is slower near the lower
  plate edge and increases with elevation.
  The plume lacks a clear origin.
  It does not match any flow topology described thus far.

  The upward-facing flow topology has bilateral symmetry; there is no
  flow between the halves created by severing along a plane of
  symmetry.  Thus, half of the upward-facing flow topology is also a
  flow topology.
  Its general formula~\eqref{eqn:general} parameters are the same as the upward-facing topology, except that~$L=2\,\Ls$.
  The $\theta=-45^\circ$ flow topology is modeled as the~$\ell^4$-norm
  of~$\overline{h'}_w$ and~$\overline{h^*}\!/2$, where
  $\overline{h^*}\!/2$~is the heat transfer from this half upward flow
  topology.  The proposed upward $\overline{h}$ formula for the 5~cm
  plate is:
$$\eqalignno{\overline{h}=k\,&\left\|{\Nuolq_w({\left|\cos\theta\right|}\,\Ra')\over L'}~,~{{\Nuols\left({\left|\sin\theta\right|}\,\Ra'/2^3\right)}\over L'/\gamma(\theta)}\right\|_4\qquad\Delta{T}\,\sin\theta\le0&\eqdef{eqn:tilt-up-5cm}\cr
  &\gamma(\theta)=\max\left(1,\min\left(2,{|\tan\theta|+1-{w/L'}}\right)\right)\qquad\qquad\quad w>L'&\eqdef{eqn:gamma}\cr}$$

  At $\theta=-90^\circ$, heat transfer is $\overline{h^*}$;
  so~$\gamma(-90^\circ)=2$. At $\theta=-45^\circ$, it is
  $\bigl\|\overline{h'}_w~,~\overline{h^*}\!/2\bigr\|_4$; so
  $\gamma(-45^\circ)=1$.
  The transition between $\gamma=2$ and $\gamma=1$ depends
  on~$\theta$, $w$, and~$L'$.  Dimensional analysis yielding
  formula~\eqref{eqn:gamma} localizes the transition to
  $w/L<\left|\tan\theta\right|<w/L+1$, whose bounds are marked by
  arrows in \figref{fig:inclined}.

\smallskip
\noindent
\vbox{\settabs 2\columns
\+\figscale{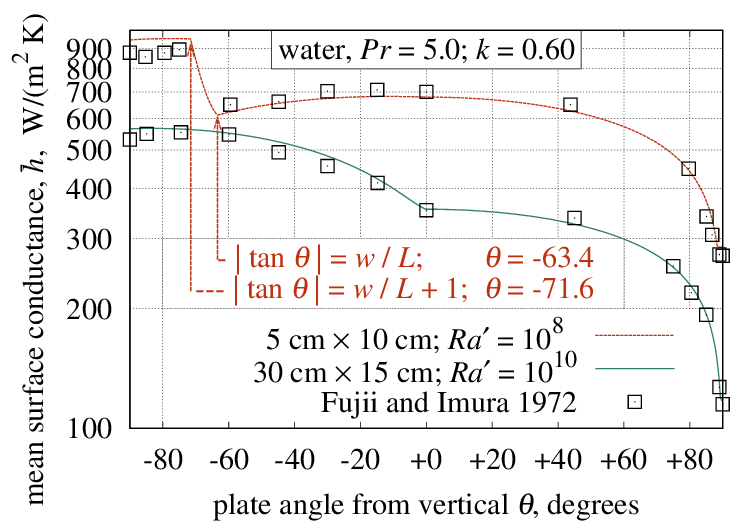}{234pt}&\hfil\figscale{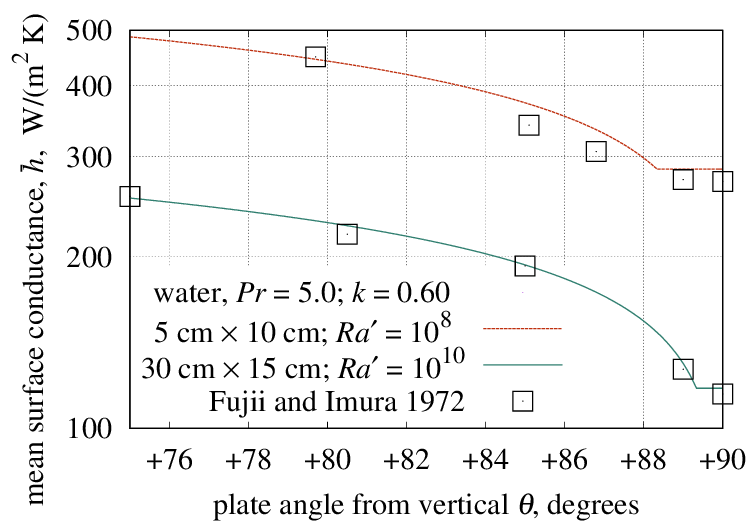}{234pt}&\cr
\+\hfil{\bf\figdef{fig:inclined}\quad inclined Fujii and Imura plates}&\hfil{\bf\figdef{fig:inclined90}\quad inclined plate detail}&\cr
}
\smallskip

\centerline{\bf\tabdef{tab:inclined-tab}\quad{inclined plate heat transfer}}
\vbox{\settabs 9\columns
\toprule
\+\hfil source&&data-set&\hfil$\Pra$&\hfil face&\quad~formula&\hfil RMSRE&\quad bias\qquad scatter&\hfill count&\cr
\midrule
\+ Fujii and Imura~\cite{fujii1972natural} -- $5~\rm{cm}\times10~\rm{cm}$&&& \hfil 5.0&\hfil inclined&\quad(\eqrefn{eqn:tilt-down},~\eqrefn{eqn:tilt-up-5cm})~$\overline{h}$&\hfill 5.8\%~\quad&\hfill$-2.4\%$\qquad& 5.3\%\hfill15&\cr
\+ Fujii and Imura~\cite{fujii1972natural} -- $30~\rm{cm}\times15~\rm{cm}$&&&\hfil 5.0&\hfil inclined&\quad(\eqrefn{eqn:tilt-down},~\eqrefn{eqn:tilt-up-30cm})~$\overline{h}$&\hfill 3.3\%~\quad&\hfill$-2.2\%$\qquad& 2.6\%\hfill14&\cr
\bottomrule
}

  \tabrefs{tab:down-conformance}, \tabrefn{tab:vertical-with-sides}, \tabrefn{tab:upward-with-sides},
  and \tabrefn{tab:inclined-tab} show that the present theory is
  sufficient to explain, with RMSRE between 2.2\% and 6.0\%, the Fujii
  and Imura heat transfer measurements of horizontal, vertical, and
  inclined plates.

\section{Results}

\centerline{\bf\tabdef{tab:conformance}\quad{measurements versus present theory}}
\vbox{\settabs 9\columns
\toprule
\+\hfil source&&data-set&\hfil$\Pra|Sc$&\hfil face&\quad~formula&\hfil RMSRE&\quad bias\qquad scatter&\hfill count&\cr
\midrule
\+ Goldstein et al~\cite{GOLDSTEIN19731025} -- sublimation&&&\hfil 2.50&\hfil                  up&\quad\eqref{eqn:upward}~$\Nuols$&\hfill  7.2\%~\quad&\hfill$-2.3\%$\qquad& 6.8\%\hfill26&\cr
\+ Fujii and Imura~\cite{fujii1972natural} -- $30~\rm{cm}\times15~\rm{cm}$&&&\hfil 5.0&\hfil   up&\quad\eqref{eqn:U-Nu_w}~$\Nuol_w$&\hfill 6.0\%~\quad&\hfill$-1.8\%$\qquad& 5.7\%\hfill11&\cr
\+ Fujii and Imura~\cite{fujii1972natural} -- $5~\rm{cm}\times10~\rm{cm}$&&& \hfil 5.0&\hfil   up&\quad\eqref{eqn:upward}~$\Nuols$&\hfill  5.0\%~\quad&\hfill$-0.4\%$\qquad& 5.0\%\hfill10&\cr
\+ Lloyd and Moran~\cite{lloyd1974natural} -- electrochemical&&&\hfil 2200&\hfil               up&\quad\eqref{eqn:upward}~$\Nuols$&\hfill  4.9\%~\quad&\hfill$+0.7\%$\qquad& 4.8\%\hfill39&\cr
\midrule
\+ Churchill and Chu~\cite{churchill1975correlating} -- Cheesewright&&&\hfil 0.70&\hfil      vertical&\quad\eqref{eqn:vertical}~$\Nuolq$&\hfill 16.4\%~\quad&\hfill$-15.4\%$\qquad& 5.6\%\hfill 6&\cr
\+ Churchill and Chu~\cite{churchill1975correlating} -- King&&&\hfil 0.70&\hfil              vertical&\quad\eqref{eqn:vertical}~$\Nuolq$&\hfill 13.5\%~\quad&\hfill $+11.1\%$\qquad& 7.6\%\hfill 8&\cr
\+ Churchill and Chu~\cite{churchill1975correlating} -- Saunders&&&\hfil 0.024&\hfil         vertical&\quad\eqref{eqn:vertical}~$\Nuolq$&\hfill  5.3\%~\quad&\hfill $-1.2\%$\qquad& 5.1\%\hfill18&\cr
\+ Fujii and Imura~\cite{fujii1972natural} -- $5~\rm{cm}\times10~\rm{cm}$&&&\hfil 5.0&\hfil  vertical&\quad\eqref{eqn:Nu'_w}~$\Nuolq_w$&\hfill   5.2\%~\quad&\hfill  $+5.0\%$\qquad& 1.6\%\hfill 6&\cr
\+ Churchill and Chu~\cite{churchill1975correlating} -- Jakob&&&\hfil 0.70&\hfil             vertical&\quad\eqref{eqn:vertical}~$\Nuolq$&\hfill  4.7\%~\quad&\hfill $+3.1\%$\qquad& 3.5\%\hfill 5&\cr
\+ Hassani and Hollands~\cite{KOBUS19953329} -- 82 mm&&&\hfil 0.71&\hfil                vertical disk&\quad\eqref{eqn:vertical}~$\Nuolq$&\hfill  3.8\%~\quad&\hfill $-3.4\%$\qquad& 1.6\%\hfill26&\cr
\+ Kobus and Wedekind~\cite{KOBUS19953329} -- three sizes&&&\hfil 0.71&\hfil            vertical disk&\quad\eqref{eqn:vertical}~$\Nuolq$&\hfill  3.2\%~\quad&\hfill $-0.4\%$\qquad& 3.1\%\hfill19&\cr
\+ Fujii and Imura~\cite{fujii1972natural} -- $30~\rm{cm}\times15~\rm{cm}$&&&\hfil 5.0&\hfil vertical&\quad\eqref{eqn:Nu'_w}~$\Nuolq_w$&\hfill   2.2\%~\quad&\hfill  $-0.5\%$\qquad& 2.2\%\hfill 5&\cr
\midrule
\+ Fujii and Imura~\cite{fujii1972natural} -- $5~\rm{cm}\times10~\rm{cm}$&&&\hfil 5.0&\hfil  down&\quad\eqref{eqn:downward}~$\Nuol_R$&\hfill  4.2\%~\quad&\hfill $-3.4\%$\qquad& 2.5\%\hfill15&\cr
\+ Aihara et al~\cite{AIHARA19722535} -- $25~\rm{cm}\times35~\rm{cm}$&&&\hfil 0.71&\hfil  down&\quad\eqref{eqn:downward}~$\Nuol_R$&\hfill  3.8\%~\quad&\hfill $-3.7\%$\qquad& 0.9\%\hfill 2&\cr
\+ Faw and Dullforce~\cite{FAW19821157} -- 18.1~cm disk&&&\hfil 0.71&\hfil  down&\quad\eqref{eqn:downward}~$\Nuol_R$&\hfill  3.7\%~\quad&\hfill $+1.8\%$\qquad& 3.2\%\hfill 3&\cr
\+ Fujii and Imura~\cite{fujii1972natural} -- $30~\rm{cm}\times15~\rm{cm}$&&&\hfil 5.0&\hfil down&\quad\eqref{eqn:downward}~$\Nuol_R$&\hfill  2.7\%~\quad&\hfill $-0.8\%$\qquad& 2.6\%\hfill 8&\cr
\midrule
\+ Fujii and Imura~\cite{fujii1972natural} -- $5~\rm{cm}\times10~\rm{cm}$&&& \hfil 5.0&\hfil inclined&\quad(\eqrefn{eqn:tilt-down},~\eqrefn{eqn:tilt-up-5cm})~$\overline{h}$&\hfill 5.8\%~\quad&\hfill$-2.4\%$\qquad& 5.3\%\hfill15&\cr
\+ Fujii and Imura~\cite{fujii1972natural} -- $30~\rm{cm}\times15~\rm{cm}$&&&\hfil 5.0&\hfil inclined&\quad(\eqrefn{eqn:tilt-down},~\eqrefn{eqn:tilt-up-30cm})~$\overline{h}$&\hfill 3.3\%~\quad&\hfill$-2.2\%$\qquad& 2.6\%\hfill14&\cr
\bottomrule
}
\smallskip

 \tabref{tab:conformance} summarizes statistics for the eighteen data-sets
 presented
 in \figrefs{fig:nuhup}, \figrefn{fig:vertplate}, \figrefn{fig:down}, \figrefn{fig:vertdisk}, \figrefn{fig:VSW}, \figrefn{fig:USW}, \figrefn{fig:inclined},
 and their associated tables.  They are grouped by orientation and
 ordered by decreasing error relative to the present work, All but
 three of these data-sets have RMSRE of~6\% or less, quantitatively
 supporting the present theory for horizontal, vertical, and inclined
 plates.

\section{Discussion}

  Renn\'o and Ingersoll~\cite{Renno.Ingersoll.JAS96} and
  Goody~\cite{doi:10.1175/1520-0469(2003)060<2827:OTMEOD>2.0.CO;2}
  found that the heat-engine efficiency limit for atmospheric
  convection is 1/2 of the reversible heat-engine efficiency limit
  $\eta$.  This investigation finds that $\eta/2$ is the limit for
  external natural convection generally.
  Reversible heat engines, such as Stirling engines, can be more
  efficient than~$\eta/2$.  External convection is not reversible.


\subsection{Side-Walls}
  The Fujii and Imura~\cite{fujii1972natural} side-walled plates do
  not qualify as external because fluid was not free to flow
  horizontally near the plate.  The side-wall formulas are not
  general, particularly for square plates.  They were investigated
  primarily to gauge how well the combination of $\ell^p$-norm with
  trigonometric scaling of~$\Ra$ explains heat transfer from inclined
  plates.

\subsection{Laminar Turbulent Transition}
  Evidence from Lloyd and Moran~\cite{lloyd1974natural}, and
  statements in Fujii and Imura~\cite{fujii1972natural} and Churchill
  and Chu~\cite{churchill1975correlating}, that the laminar-turbulent
  transition was irrelevant to natural convection heat transfer were
  published in the early 1970s.  Yet, belief that it governs
  external plate natural convection heat transfer has
  persisted~\cite{goldstein_lau_1983,rohsenow1998handbook,KITAMURA2015320}.

  Much of the subsequent natural convection literature investigates
  local flow properties.  Such studies do not inform this
  investigation's systemic-invariant analysis.  However, streamlines
  photographs were crucial to characterizing the flow topologies and
  deriving the present formulas.

\subsection{Pause in Horizontal Flow}
  The pause in horizontal flow above the upward-facing 5~cm plate is
  visible in the $\theta=-90^\circ$ photograph from Fujii and
  Imura~\cite{fujii1972natural}.  Their photographs of vertical and
  downward-facing plate streamlines do not include enough of the fluid
  above the plates to see the horizontal pause expected by this
  investigation.



\subsection{Harmonic Mean}
  The harmonic mean integral in formulas~\eqref{eqn:L_R} and~\eqref{eqn:L'}
  converges only when the perimeter curve is perpendicular to the
  integration axis at both integration limits.
  For downward-facing plates this includes all circles, ellipses, and
  rectangles; for vertical plates this includes all circles and ellipses,
  and trapezoids (including rectangles) with two vertical edges.
  When~$\root3\of{\Ra'}\gg1$, vertical plate $\Nuolq$ scales
  with~$L'$.  Hence, the heat flow rate $\overline{h'}=k\,\Nuolq/L'$
  is sensitive to~$L'$ only at small~$\Ra'$ values.




\section{Conclusions}

  Using a novel methodology based on streamline photographs,
  dimensional analysis, and the thermodynamic constraints on
  heat-engine efficiency, this investigation:

\unorderedlist
 \li derived from first principles a previously unknown comprehensive
  heat transfer formula for upward-facing convection;

 \li extended the scope of vertical and downward-facing formulas from
  rectangles to other shapes with convex perimeters using the harmonic
  mean for the characteristic-length metric; and

 \li unified the $\Pra$ dependence of vertical and downward-facing
  plates.
\endunorderedlist

  A comprehensive, exact model for natural convection from an
  external, isothermal flat surface can now be succinctly stated.  For
  horizontal upward-facing plates:

$$\eqalign{
  \Nuols(\Ra^*)&=\Nuzs\,\left\|1-{1\over\sqrt{8}}~,~{\Nuzs^{1/3}\over4}\,\root3\of{\Ra^*}\right\|_{1/2}
  \qquad\Ra^*>1\cr
  \left\|F_0~,~F_1\right\|_p&=\left(|F_0|^p+|F_1|^p\right)^{1/p}
  \qquad\Nuzs={2\over\pi}\approx0.637\cr}$$

  For vertical and downward-facing plates, the convection reduction
  due to self-obstruction is the $\Ra$ scaling factor $1/\,\Xi(\Pra)$,
  yielding comprehensive formulas for vertical $\Nuolq$ and
  downward~$\Nuol_R$:
$$\eqalign{
  \Nuolq(\Ra')&=\left\|{\Nuzq\over2}~,~{\Nuzq^{4/3}\over8\,\root3\of2}\left[{\Ra'\over\Xi(\Pra)}\right]^{1/3}\right\|_{1/2}\qquad\Ra'>1\cr
  \Nuol_R(\Ra_R)&=~\,{\Nuzq\over4}+{\Nuzq^{6/5}\over2^{7/5}}\left[{\Ra_R\over\Xi(\Pra)}\right]^{1/5}\quad\quad\quad~\Ra_R>1\cr
  \qquad\Xi(\Pra)&=\left\|1~,~{0.5\over\Pra}\right\|_{\sqrt{1/3}}\qquad\Nuzq={8^{5/4}\over\pi^2}\approx1.363\cr}$$

  The reduction in effective gravitational acceleration $g$ also
  scales~$\Ra$.  The average convective surface conductance for an
  external, isothermal plate inclined at angle $\theta$ from vertical
  is:
$$\overline{h}=k\cases{
  \max\left({\Nuolq({\left|\cos\theta\right|}\,\Ra')/L'}, {\Nuols\left({\left|\sin\theta\right|}\,\Ra'\,[{\Ls/L'}]^3\right)/\Ls}\right)&$\Ra'\sin\theta<-[\Ls/L']^3$\cr
  \max\left({\Nuolq({\left|\cos\theta\right|}\,\Ra')/L'}, {\Nuol_R\left({\left|\sin\theta\right|}\,\Ra'\,[{L_R/L'}]^3\right)/L_R}\right)&$\Ra'\sin\theta>[L_R/L']^3$\cr
  {\Nuolq\left({\left|\cos\theta\right|}\,\Ra'\right)/L'}&otherwise.\cr
}$$

\unorderedlist

 \li The upward-facing characteristic-length $\Ls$ is the area-to-perimeter
 ratio.

 \li The vertical characteristic-length $L'$ is the harmonic mean of
 the perimeter vertical spans.

 \li The downward-facing characteristic-length $L_R$ is the harmonic
 mean of the perimeter distances to that bisector which is
 perpendicular to the shortest bisector.

 \li $\Ra'$ is the Rayleigh number computed with vertical
 characteristic-length $L'$.

\endunorderedlist

 The harmonic mean metrics extend vertical~$\Nuolq$ and
 downward-facing~$\Nuol_R$ to non-rectangular plates.

\medskip


  The present theory was compared with eighteen data-sets from seven
  peer-reviewed articles, testing circular, rectangular, and inclined
  rectangular plates, laminar and turbulent flows, with
  $0.024<\Pra<2200$ and $1<\Ra<10^{12}$.  All except three of the
  data-sets had between 2\% and 6\% RMSRE from the present theory.

\medskip

\unorderedlist

 \li The $\Nuols$ formula improves accuracy and $\Ra$ range
  substantially over the piece-wise power-laws currently employed for
  predicting upward convection heat transfer.

 \li With less than 1\% difference between $\Nuolq$ and the Churchill
  and Chu (1975) vertical formula, there is little need to replace it
  in existing applications.

 \li However, the published Schulenberg (1985) formula can return values
   which are 10\% smaller than $\Nuol_R$; $\Nuol_R$ should replace it.


\endunorderedlist



\section{Nomenclature}

\nomenclature[A]{$A$}{plate area ($\rm m^2$)}
\nomenclature[A]{$c_p$}{fluid specific heat at constant pressure (${\rm J/(kg\cdot K)}$)}
\nomenclature[A]{$B$}{sum of mean lengths of plate parallel flows divided by $L$}
\nomenclature[A]{$C$}{plate area fraction responsible for flow induced heat transfer}
\nomenclature[A]{$D$}{effective length of heat transfer contact with plate divided by $L$}
\nomenclature[A]{$E$}{count of $90^\circ$ changes in direction of fluid flow}
\nomenclature[A]{$g$}{gravitational acceleration (${\rm m/s^2}$)}
\nomenclature[A]{$\overline{h}$}{average convective surface conductance (${\rm W/(m^2\cdot K)}$)}
\nomenclature[A]{$I$}{power flux (${\rm W/m^2}$)}
\nomenclature[A]{$k$}{fluid thermal conductivity (${\rm W/(m\cdot K)}$)}
\nomenclature[A]{$L$}{characteristic-length (m)}
\nomenclature[A]{$M$}{air molar mass (${\rm kg}$)}
\nomenclature[A]{$\Nuol$}{average Nusselt number}
\nomenclature[A]{$P$}{air pressure (${\rm N/m^2}$)}
\nomenclature[A]{$\Pra$}{Prandtl number}
\nomenclature[A]{$q$}{conduction power (W)}
\nomenclature[A]{$q_{SS}^*$}{dimensionless conduction shape factor}
\nomenclature[A]{$R$}{disk radius (m)}
\nomenclature[A]{$\overline{R}$}{universal gas constant (${\rm J/(kg\cdot K)}$)}
\nomenclature[A]{$\Ra$}{Rayleigh number}
\nomenclature[A]{$\Rey$}{Reynolds number}
\nomenclature[A]{$Sc$}{Schmidt number}
\nomenclature[A]{$S$}{conduction shape factor (m)}
\nomenclature[A]{$T$}{temperature (K)}
\nomenclature[A]{$u$}{fluid velocity ($\rm m/s$)}
\nomenclature[A]{$V$}{air volume ($\rm m^3$)}
\nomenclature[A]{$W$}{work (J)}
\nomenclature[A]{$w$}{distance between side-walls (m)}
\nomenclature[A]{$y_+,~y_-$}{perimeter functions (m)}
\nomenclature[A]{$\overline{y}$}{Harmonic mean of perimeter functions (m)}

\noindent{\bf Greek Symbols}

\nomenclature[G]{$\Delta{Q}$}{heat energy (J)}
\nomenclature[G]{$\Delta{T}$}{temperature difference $=T-T_\infty$ (K)}
\nomenclature[G]{$\alpha$}{fluid thermal diffusivity $=k/[\rho\,c_p]$ (${\rm m^2/s}$)}
\nomenclature[G]{$\beta$}{fluid thermal expansion coefficient ($\rm K^{-1}$)}
\nomenclature[G]{$\eta$}{thermodynamic heat-engine efficiency}
\nomenclature[G]{$\nu$}{fluid kinematic viscosity (${\rm m^2/s}$)}
\nomenclature[G]{$\Pi$}{dimensionless variable group}
\nomenclature[G]{$\Phi$}{power flux (${\rm W/m^2}$)}
\nomenclature[G]{$\rho$}{fluid density (${\rm kg/m^3}$)}
\nomenclature[G]{$\theta$}{surface angle from vertical ($-90^\circ$ is face up)}
\nomenclature[G]{$\Xi$}{$\Ra$ self-obstruction (reciprocal) factor}

\noindent{\bf Superscripts}

\nomenclature[X]{$*$}{upward-facing plate}
\nomenclature[X]{$'$}{vertical plate}

\noindent{\bf Subscripts}

\nomenclature[Z]{$0$}{conduction}
\nomenclature[Z]{$\forall$}{unified}
\nomenclature[Z]{$A$}{atmospheric natural convective}
\nomenclature[Z]{$h$}{heated}
\nomenclature[Z]{$i$}{induced flow along plate}
\nomenclature[Z]{$k$}{kinetic}
\nomenclature[Z]{$N$}{natural convective}
\nomenclature[Z]{$p$}{plate}
\nomenclature[Z]{$R$}{downward-facing plate}
\nomenclature[Z]{$r$}{downward-facing disk}
\nomenclature[Z]{$S$}{dimensionless shape factor}
\nomenclature[Z]{$t$}{ideal turbine}
\nomenclature[Z]{$w$}{with side-walls}
\nomenclature[Z]{$\infty$}{bulk fluid}
\nomenclature[Z]{$?$}{possibly misprinted formula}


\beginsection{Acknowledgments}

Thanks to Dave Custer, Rich Hilliard, Roberta Jaffer, and anonymous
reviewers for their useful suggestions.

\section{References}

\bibliographystyle{unsrtDOI}
\bibliography{citations}

\vfill\eject
\bye